\documentclass[a4paper,useAMS,usenatbib]{mnras}
\usepackage{amsmath}
\usepackage{amssymb}
\usepackage{float}
\usepackage{graphicx}
\usepackage{hyperref}
\usepackage{subcaption}
\usepackage{epstopdf}
\usepackage{array}
\usepackage{color}
\title{Precision cosmology with baryons: non-radiative hydrodynamics of galaxy groups}

\author[Various authors]{Manuel Rabold $^{1,2,}$\thanks{manuel@physik.uzh.ch} and Romain Teyssier$^{1,2}$ \\
$^1${Center for Theoretical Astrophysics and Cosmology, University of Zurich, Winterthurerstrasse 190, 8057 Zurich, Switzerland} \\
$^2${Institute for Computational Science, University of Zurich, Winterthurerstrasse 190, 8057 Zurich, Switzerland} \\
}
\begin{document}
\maketitle

\begin{abstract}
The effect of baryons on the matter power spectrum is likely to have an observable effect for future galaxy surveys, like Euclid or LSST. 
As a first step towards a fully predictive theory, we investigate the effect of non-radiative hydrodynamics on the structure of galaxy groups sized halos, which contribute the most to the weak lensing power spectrum. 
We perform high resolution (more than one million particles per halo and one kilo-parsec resolution) non-radiative hydrodynamical zoom-in simulations of a sample of 16 halos, comparing the profiles to popular analytical models.
We find that the total mass profile is well fitted by a Navarro, Frenk \& White model, with parameters slightly modified from the dark matter only simulation. 
We also find that the Komatsu \& Seljak hydrostatic solution provides a good fit to the gas profiles, 
with however significant deviations, arising from strong turbulent mixing in the core and from non-thermal, turbulent pressure support in the outskirts. 
The turbulent energy follows a shallow, rising linear profile with radius, and correlates with the halo formation time.
Using only three main structural halo parameters as variables (total mass, concentration parameter and central gas density), 
we can predict with an accuracy better than 20\% the individual gas density and temperature profiles. 
For the average total mass profile, which is relevant for power spectrum calculations, we even reach an accuracy of 1\%.
The robustness of these predictions has been tested against resolution effects, different types of initial conditions and hydrodynamical schemes.
\end{abstract}


\section{Introduction}
\label{introduction}

Our decade has seen Cosmic Microwave Background (CMB) experiments dominate the field of observational cosmology, allowing us to determine cosmological parameters with unprecedented
accuracy \citep{2011ApJS..192...18K,2016A&A...594A..13P}. The next decade will be the era of precision cosmology based on large scale galaxy surveys, like Euclid \citep{2011arXiv1110.3193L} and LSST \citep{2012arXiv1211.0310L}. In order for cosmological probes such as Weak Lensing (WL) or Galaxy Clustering (GC) to be really competitive, when compared to CMB data, it is required to achieve sub-percent level accuracy in the determination of the matter power spectrum, at scales of comoving wavelength $0.1\,h\,\mathrm{Mpc}^{-1} < k < 10\,h\,\mathrm{Mpc}^{-1}$ \citep{2005APh....23..369H}. This brings considerable challenges to the instrumental design and to the data analysis of these future experiments, especially regarding the understanding and the control of systematics errors. 

An important source of these systematic errors is coming from the theoretical predictions for both GC and WL probes. Although the matter distribution is well understood at the linear level, below modes of $0.1\,h\,\mathrm{Mpc}^{-1}$, this is not necessarily the case on smaller scales where non-linear effects become important. Collisionless N-body simulations, on one hand, have been successful in modelling dark matter only universes, with an accuracy better than $1\%$ at $k < 3\,h\,\mathrm{Mpc}^{-1}$ and better than $3\%$ at $k < 10\,h\,\mathrm{Mpc}^{-1}$ \citep{2016JCAP...04..047S}. 
Baryonic effects, on the other hand, cannot be ignored on these scales: in the range of $1\,h\,\mathrm{Mpc}^{-1} < k < 10\,h\,\mathrm{Mpc}^{-1}$, baryonic physics can modify the total matter power spectrum (compared to the naive dark matter only case) up to 10\% \citep{2011MNRAS.415.3649V,2015JCAP...12..049S}, considering pessimistic but plausible scenarios.

To account for baryons in a cosmological simulation, we model them as a collisional fluid. As a first basic step, this gas component can be considered as purely non-radiative. 
From there, one can introduce additional physical processes to the baryonic gas, like radiative cooling, star formation, supernova feedback or AGN feedback 
\citep[see][for a review]{2008SSRv..134..229D}. The latter, though still in an experimental, not really predictive phase, has proven very valuable in exploring strong baryonic effects on large scale, and in reproducing many observable properties of large galaxy clusters and groups. The ultimate goal for the inclusion of baryonic effects into the analysis of our cosmological datasets, would be to parameterise their influence onto the relevant spectra, in the form of one or a few key parameters, in addition to the set of standard cosmological parameters, so that we could fit those parameters together with the important dark sector parameters.

One of the advantages coming from the utilisation of the cosmological weak lensing technique, is its sensitivity on the late time evolution of the universe during the dark energy dominated era. So that a particular high expectation lies on the determination of the parameters relevant herein, like the dark energy equation of state parameter and the neutrino masses. Further, the corresponding redshift range relevant for cosmological weak lensing translates, through the halo mass function, into the result that the strongest signal will come from galaxy group size halos \citep{2000MNRAS.318..203S}. For this reason our analysis focuses on the latter.

Since galaxy clusters are the standard testbeds for cosmological studies \citep{2011ASL.....4..204B, 2012ARA&A..50..353K, 2011ARA&A..49..409A}, especially for the comparison of numerical predictions with observational results, galaxy group size halos, have played a subdominant role in the analysis of cosmological structure, mostly in combination with studies on cluster size halos. Indeed, from a naiv view point, galaxy group halos are scaled down versions of galaxy cluster, depending on the characteristic length scales that are involved in the implemented physical processes. For purely non-radiative hydrodynamics, no characteristic length scales exists, since both the gravitational and hydrodynamical equations of motion are scale free. On second thought however, the difference in size between group and cluster halos, translates into a different time scale for their gravitational collapse. So that one could expect group size halos to be ahead of their cluster sized counterparts, in the collapse process, and to be more concentrated accordingly.

Our work is to our knowledge the first precision cosmology numerical study focusing entirely on  group size halos. Previous work has been done in the area of galaxy formation physics within  groups \citep{2010ApJ...709..218F} and on testing of new implementations of AGN feedback \citep{2010MNRAS.406..822M}. This, however, is not the focus of this paper.    
The analysis of this paper concentrates on the first basic step to include baryons: we investigate their influence as they are modelled as a purely non-radiative gas. Our aim is also to quantify the effect of numerical resolution, which is needed to resolve the inner regions of halos $r<10~\mathrm{kpc}$ with the required precision. As a reference, we will use a well-known analytical model for the radial dependence of the thermodynamic quantities of the baryonic component, based on hydrostatic equilibrium and a polytropic Equation-of-State, from \cite{2001MNRAS.327.1353K}. 

A similar study about non-radiative hydro simulations of cluster and group size halos was reported in \cite{2003MNRAS.346..731A}. There however the focus lay more on the testing of analytic models for the radial dependence of the thermodynamical properties of the baryonic component, commonly named the intracluster medium (ICM) and the intragroup medium (IGM). The ICM is an important source of information for observations of galaxy clusters, since it consist largely of ionized hydrogen, which is continuously emitting X-ray photons. The spectrum of these photons contains information, about the spatial distribution and thermodynamical state of the emitting baryonic gas.

Other important numerical studies of ICM/IGM profiles with non-radiative hydro simulations were done by \cite{1995MNRAS.275..720N}, \cite{1998ApJ...503..569E}, \cite{2002ApJ...579..571L},\cite{2004MNRAS.351..237R}, \cite{2006MNRAS.373.1339R} and \cite{2007ApJ...668....1N}.

Aside from the intrinsic thermodynamical properties of the baryonic gas, we are also addressing its property of a non-thermal pressure component arising from turbulent motion in the gas. The question about how this additional pressure support enters into the assumption of hydrostatic equilibrium has been addressed before, in the studies of \cite{2004MNRAS.351..237R}, \cite{2006MNRAS.369.2013R}, \cite{2006MNRAS.373.1339R}, \cite{2009ApJ...705.1129L}, \cite{2008A&A...491...71P}. The importance of this issue comes from the fact that it leads to a mismatch in the mass estimate by X-ray observations of galaxy groups and clusters when the equation of hydrostatic equilibrium (HSE) is applied, because there only the thermal pressure of the X-ray emitting gas is accounted for \citep{2011MNRAS.413..573B}. However, since the non-thermal component is more accessible in simulations, analytical fits have been proposed for its radial dependence. Recent works are \cite{2016ApJ...827..112B}, \cite{2016arXiv160804388M} and the paper series \cite{2014MNRAS.442..521S} \cite{2015MNRAS.448.1020S}, \cite{2016MNRAS.455.2936S}.  

Throughout the evolution of numerical astrophysical modelling, the quantification of numerical effects onto the simulation results has a long tradition. Already for the stage of pure N-body simulations it is important to test for numerical convergence as was done by the studies of \cite{2003MNRAS.338...14P} and \cite{2008CS&D....1a5003H}. Also for non-radiative hydro simulations a series of code comparison projects were performed early on by \cite{1994ApJ...430...83K}, \cite{1999ApJ...525..554F} and \cite{2005ApJS..160....1O}, and most recently in the nIFTy project \citep{2016MNRAS.457.4063S}. An established method here is to give a fixed set of initial conditions to each of the participating code developers, for each to run their simulation with their own choice of numerical parameters. 

One important feature in these analyses are halo profiles and maps, which provide a common ground for the comparison of results from different numerical setups or codes. In the N-body case for density and mass and in the case where hydrodynamics is included also for thermodynamical quantities of the baryonic component like temperature, pressure or entropy. Our project focuses on halo profiles and maps alone. Further quantities used for comparisons could be the halo mass function, or the matter power spectrum itself.

Key questions that we address are the followings.
\begin{itemize}
\item how strong are the deviations of the numerical results from the analytic model, on average and for individual halos?
\item how strong is the deviation from the numerical mean coming from the individual nature of the halos? In other words: how strong is the scatter?
\item how strongly do numerical parameters e.g. the maximum resolution, or the initial conditions influence the results? 
\item to which inner radius can the simulations be considered numerically converged?
\item are state of the art computer simulations capable of reaching the precision required by future observation?
\end{itemize}

Since these technical/numerical/systematic questions are the focus of this paper, we carry our analysis out for the simplest method of cosmological simulations, incorporating baryons: collisionless dark matter particles plus a non-radiative baryonic fluid. Parameter studies of cosmological hydrodynamic simulations including radiative processes, star formation, stellar/AGN feedback and further subgrid physics will be presented in subsequent publications.

This paper is structured as follows: Section \ref{simulation_parameters} describes our numerical simulations. Section \ref{analytical_Model} describes the analytical model to which we compare our findings. In Section \ref{halo_selection} the selection criteria for our halo catalog are discussed. In Section \ref{results} we present the results in form of halo profiles of the relevant physical quantities and in Section \ref{correlations} correlations between the halo properties are depicted. The investigation of numerical effects is described in Section \ref{effect_of_numerical_parameters}, while Section \ref{conclusions} concludes the paper.

\section{Simulation Parameters}
\label{simulation_parameters}

All simulations were performed with the Adaptive Mesh Refinement code RAMSES \citep{2002A&A...385..337T} in a cosmological periodic cubic box of side length $L=100\,\rm{Mpc}\,h^{-1}$. Initially a reference run with dark matter only needed to be done, so that the halo regions of interest could be selected from it (see the particular section for the identification). Simulations of 16 individual halos were undertaken with the use of the zoom-in technique. This allows to refine the region of interest (in our case the $2\,r_{200}$ environment of each halo) with a higher resolution, than the rest of the box, while leaving the box size constant. The underlying cosmological model is characterized by the parameters listed in Table~\ref{Cosmo_Params}. Initial conditions for all simulations were provided by the MUSIC code \citep{2011MNRAS.415.2101H}. For the definition of $r_{200}$ the average matter density of the universe $\bar{\rho}_m$ is used.

\begin{table}
\centering 
\begin{tabular}{c  c}
\\
\hline
$\Omega_{b,0}$ 		& 0.04825 \\
$\Omega_{m,0}$ 		& 0.308 \\
$\Omega_{\Lambda,0}$ 	& 0.692 \\
$h_0$ 			& 0.6777 \\
$n_s$ 			& 0.9611 \\
$\sigma_8$	 	& 0.8288 \\
$w$ 			& -1\\
\hline
\end{tabular}
\caption[Adopted cosmological parameters]{Adopted cosmological parameters}
\label{Cosmo_Params}
\end{table} 

The effective size of the reference run`s initial grid was $512^3$ ($\ell_{\rm min}=9$). This translates into a mass resolution of $m_{DM}=9.34 \times 10^8 \,M_{\odot}$. During the run, 7 more levels were added recursively. Once a halo of interest is found at $z=0$, a bounding sphere of twice the halos virial radius around the halos centre of mass is set up, to account for all particles within it. The same particles are then identified in the initial grid at $z=100$.  A bounding ellipsoid of minimum size including all relevant particles is generated. Its geometric information is passed on to the MUSIC code, which in turn creates the multi-level initial conditions for the zoom-in run.

For the zoom-in runs, the levels of the initial grid ran from $\ell_{\rm ini}=11$ in the region of interest, down to $\ell_{\rm min}=7$ in the rest of the box. $\ell_{\rm ini}=11$ corresponds to a mass resolution $m_{DM}=1.46 \times 10^7 \,M_{\odot}$ in dark matter only case, and to $m_{DM}=1.25 \times 10^7 \,M_{\odot}$ and $m_{b}=2.13 \times 10^6 \,M_{\odot}$ in the runs including hydrodynamics. The dynamical refinement is implemented in the following way (quasi-Lagrangian): when the dark matter mass or the baryon mass in a cell reaches eight times the initial mass resolution, during the run, the cell is split into 8 children cells. This results in a nearly constant spatial resolution in physical units throughout the run. The chosen maximum refinement of level $\ell_{\rm max}=18$ corresponds to a physical minimum cell size of $\Delta x_{\rm min}=L/2^{\ell_{\rm max}}=1.1258\,\mathrm{kpc}$. 
We use for the  hydro solver a second-order unsplit Godunov scheme based on the HLLC Riemann solver \citep{2006JCoPh.218...44T} and the MinMod slope type limiter \citep{2006A&A...457..371F}.

\section{Halo Finding}
\label{halo_finding}

For halo detection, we applied the HOP halo finder \citep{1998ApJ...498..137E, 2010ApJS..191...43S} to the dark matter particles in our simulations.
HOP calculates the density of each particle, from a specified number of its nearest neighbours.
With this information it assigns each particle to a local density peak (densest neighbour), which is found after checking another specified number of neighbouring particles.
Now all particles are assigned to a group defined by its densest particle, or they themselves are the densest of a number of particles (the group). 
In the next step particles whose density is below a specified density contrast $\delta_{outer}$ are removed from these groups. 
This way it is decided, which particles belong to a halo and which do not.  
On top of that, the issue has to be addressed, that a dens spatial region could contain more than one density maximum. 
This means that such a region, which would in the physical sense correspond to a halo, is artificially split into smaller groups, defined by their local density maxima, when the aforementioned steps are applied. 
To overcome this mishap, the groups found so far are merged together in another step, which introduces two additional density contrast parameters $\delta_{saddle}$ and $\delta_{peak}$. Only a group, whose highest density lies above $\delta_{peak}$ can be an alone standing halo, otherwise it is merged into another halo, which is defined by a group, whose highest density exceeds $\delta_{peak}$. 
The question, to which halo it should be merged is decided by, with which halo it shares the boundary of highest density.
Further, two neighbouring groups whose highest densities both exceed $\delta_{peak}$ are merged together, if the density at their boundary exceeds the value $\delta_{saddle}$. So typically one has $\delta_{outer} < \delta_{saddle} < \delta_{peak}$.
The most important parameter is $\delta_{outer}$. 
We have selected the following set of density thresholds: $(\delta_{outer},\,\delta_{saddle},\,\delta_{peak})=(80,200,240)$. 
The resulting halo definition corresponds to the one of using a Friends-of-Friends halo finder, with linking length 0.2 Mpc \citep{1998ApJ...498..137E}. 
From the latter in turn, it is known that the resulting halo mass is a good estimate for $M_{200}$, defined with the average mass density of the universe.
Since the implementation of HOP used in this work, does not output any information about substructure, we have not removed any particles from the found halos in the following, to unbind subhalos.


\section{The analytic Model}
\label{analytical_Model}

The analytical model to which we compare our simulation data is based on the following principles.
The density radial profile of the halos overall matter distribution follows the NFW model \citep{1996ApJ...462..563N}:
\begin{equation}
\rho_{tot}(r)=\frac{\rho_{s}}{\frac{r}{r_{s}}\left({1+\frac{r}{r_{s}}}\right)^2}
\end{equation}
This model introduces two parameters $\rho_s$ and $r_s$.
The temperature and density radial profiles of the baryonic fraction follow the \cite{2001MNRAS.327.1353K} analytical hydrostatic and polytropic model:      
\begin{equation}
T_{gas}(r)=T_{0}\frac{\ln(1+\frac{r}{r_{s}})}{\frac{r}{r_{s}}}
\label{Temp}
\end{equation}
\begin{equation}
\rho_{gas}(r)=\rho_{0}\left({\frac{\ln(1+\frac{r}{r_{s}})}{\frac{r}{r_{s}}}}\right)^{\frac{1}{\Gamma-1}}
\label{Gas_dens}
\end{equation}
where $\Gamma$ is the adopted polytropic index
used to represent the polytropic equation of state
\begin{equation}
P_{gas}(r)\propto \rho_{gas}(r)^{\Gamma}
\end{equation}
$\Gamma$ together with $T_0$ and $\rho_0$ add three additional parameters to the model. $T_0$ the normalisation of the temperature profile is however determined already by the condition of zero pressure at infinity, which results in
\begin{equation}
T_0=\frac{4 \pi G \rho_{s} r^2_{s} \mu m_p }{k_B}\frac{\Gamma-1}{\Gamma}
\label{T_0}
\end{equation}
Let us explain how we determine the values of the various parameters for each halo. First, $\rho_s$  and $r_s$ can be extracted for each of the 16 halos by fitting to their circular velocity curve. 
This is done by making a least squares fit to the circular velocity squared curve: 
\begin{equation}
L(r_s,\rho_s)=\sum_{i=1}^{N}\left({V^2_i-F^2(r_i)}\right)^2 
\end{equation}
Here $r_i$ and $V_i$ are data points of our halo profiles, and $F(r_i)$ is the circular velocity for the NFW case:
\begin{equation}
F^2(r)=4 \pi \,G\, \rho_s \, r_s^3 \frac{1}{r} \cdot \left[{\ln(1+\frac{r}{r_{s}})-\frac{r/r_s}{1+r/r_s}}\right]
\end{equation}
We give each data point pair an equal weight. 
The intervall which we consider reaches from the resolution limit $\Delta x_{\rm min}=1.1258\,\mathrm{kpc}$ to $1.5\,r_{HOP}$ (for the definition see Appendix \ref{halo_mass_definitions}). 
This corresponds to between 550 and 600 data points, depending on the size of the halo. 
We have also tried the range $[1.1258\,\mathrm{kpc};r_{HOP}]$ for comparison, but this affected the resulting values for $r_s$ and $\rho_s$ only insignificantly.

\begin{figure}
\centering
\includegraphics[scale=0.575]{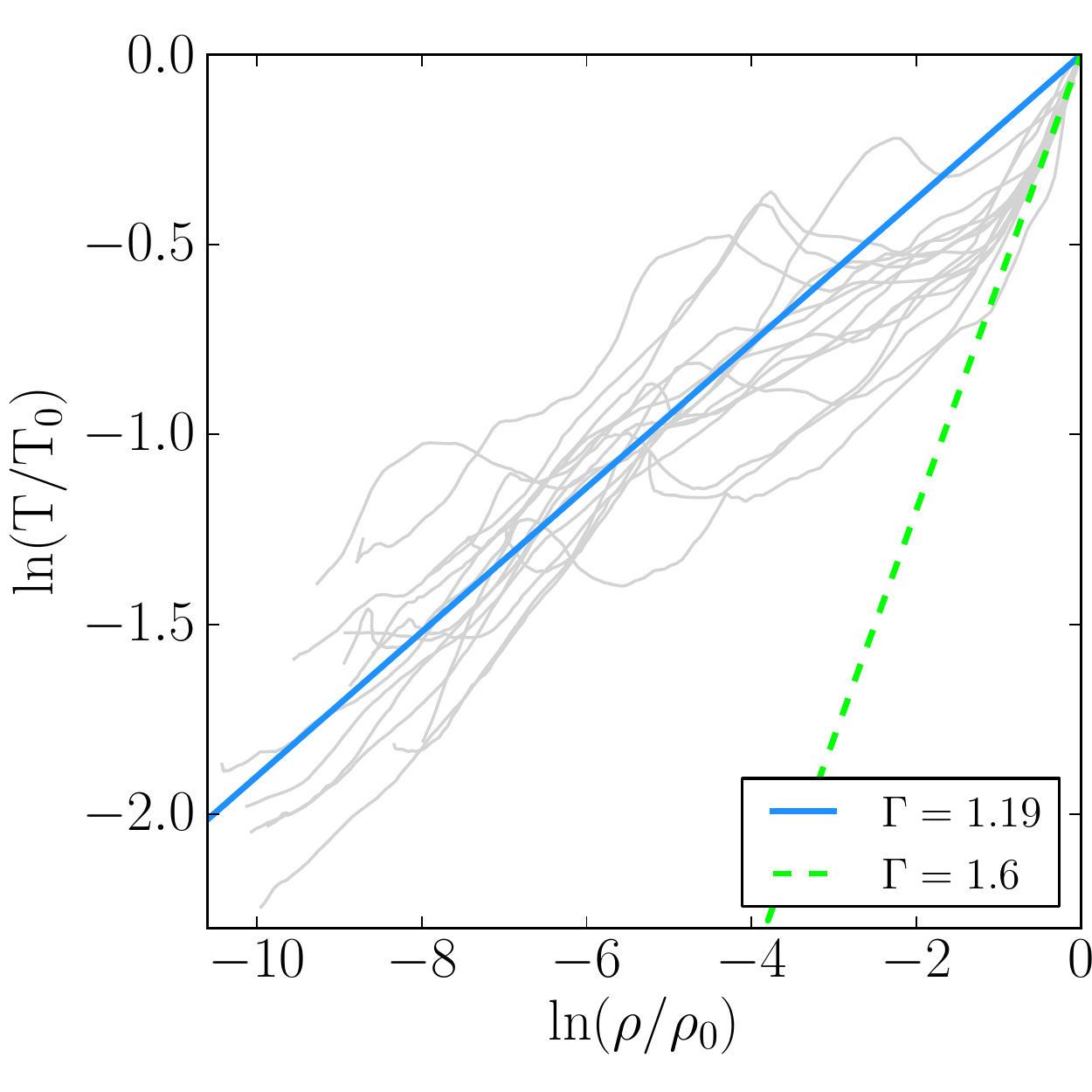}
\caption{Logarithm of the temperature profile versus the logarithm of the density profile for all 16 halos in our sample. For comparison, we show the polytropic relation with $\Gamma=1.19$ as a solid line and another polytropic relation with $\Gamma=1.6$, more adapted to the central regions of our sample.}
\label{gamma}
\end{figure}

The next parameter $T_0$ follows directly from Equation~(\ref{T_0}). We have  explored various values for $\Gamma$ with  $1.15<\Gamma<1.21$. The value which we found to fit our numerical density and temperature curves the best is $\Gamma=1.19$, very close to the value $\Gamma=1.18$ suggested by \cite{2003MNRAS.346..731A}. To confirm this, we plot the logarithm of the temperature versus the logarithm of the density for the radial profiles of all 16 halos in Figure~\ref{gamma}. Note that in the very inner part, our simulations prefer a steeper value with $\Gamma=1.6$, but our adopted value is better at reproducing the density/temperature relation over the entire range of densities. Finally, the gas profile normalisation $\rho_0$ is fixed by assuming the additional constraint
\begin{equation}
M_{gas}(r_{200})=\frac{\Omega_b}{\Omega_m}M_{tot}(r_{200}),
\label{normalization}
\end{equation}
which translates into
\begin{equation}
\rho_0=\frac{\Omega_b}{\Omega_m}\rho_s \frac{\ln{(1+c)}-\frac{c}{1+c}}{\int_0^c \left( \frac{\ln{(1+x)}}{x} \right)^{\Gamma}x^2 {\rm d}x},
\label{normalization}
\end{equation}
which, for our adopted cosmology and for the whole range of interest $4<c<20$, translates into a simple approximation, accurate to $\pm~2\%$,
\begin{equation}
\rho_0 \simeq 0.208 \rho_s.
\end{equation}
Note that $r_{200}$ is determined for each halo through the function 
\begin{equation}
\Delta(r) = \frac{3}{4\pi}\frac{M(<r)}{\bar{\rho}_m r^3}.
\label{Delta200} 
\end{equation}
$r_{200}$ is simply the radius for which $\Delta(r)=200$ in units of the average density in the universe, and its value is fully fixed once the 2 parameters $\rho_s$ and $r_s$ have been found.\\
The halo mass $M_{200}$ directly follows.

\section{Halo Selection}
\label{halo_selection}

We identified galaxy group sized halos in the N-body unigrid simulation at $z=0$. 
At this redshift, the simulation volume contained 73'947 halos with masses in the mass range of interest $ 5 \times 10^{12} M_{\odot} h^{-1} < M <  5 \times 10^{13} M_{\odot} h^{-1}$. 
Please note that these are the masses given by the halo finder HOP. The masses of our selected halos, listed in Table \ref{Halo_Table} as $M_{200}$, are however the ones which come out of the NFW model, once the halo parameters $r_s$ and $\rho_s$ have been derived. 
In Appendix \ref{halo_mass_definitions} we give all details about the different methods to obtain the halo masses, and hence halo radii $r_{200}$. 
For the remaining part of this paper, we always refer to the mass and radii given by the NFW model, with $M_{200}$ and $r_{200}$. 
To the 73'947 found halos, we applied a series of selection criteria to ensure that we have a representative sample of group size halos, considering their mass accretion history and their circular velocity curves (as a measure of their spatial mass distribution). 
Finally, we ensured that the entire mass range is represented, by selecting quasi-randomly 16 halos with masses distributed over the entire interval. 
Our selection is not entirely random, as we preferentially select relatively isolated halos, since isolated halos are easier to handle in zoom-in simulations. 
For a detailed description of the selection process, we refer to Appendix \ref{halo_selection_details}. 
 
\begin{figure*}
\centering
\includegraphics[scale=0.7]{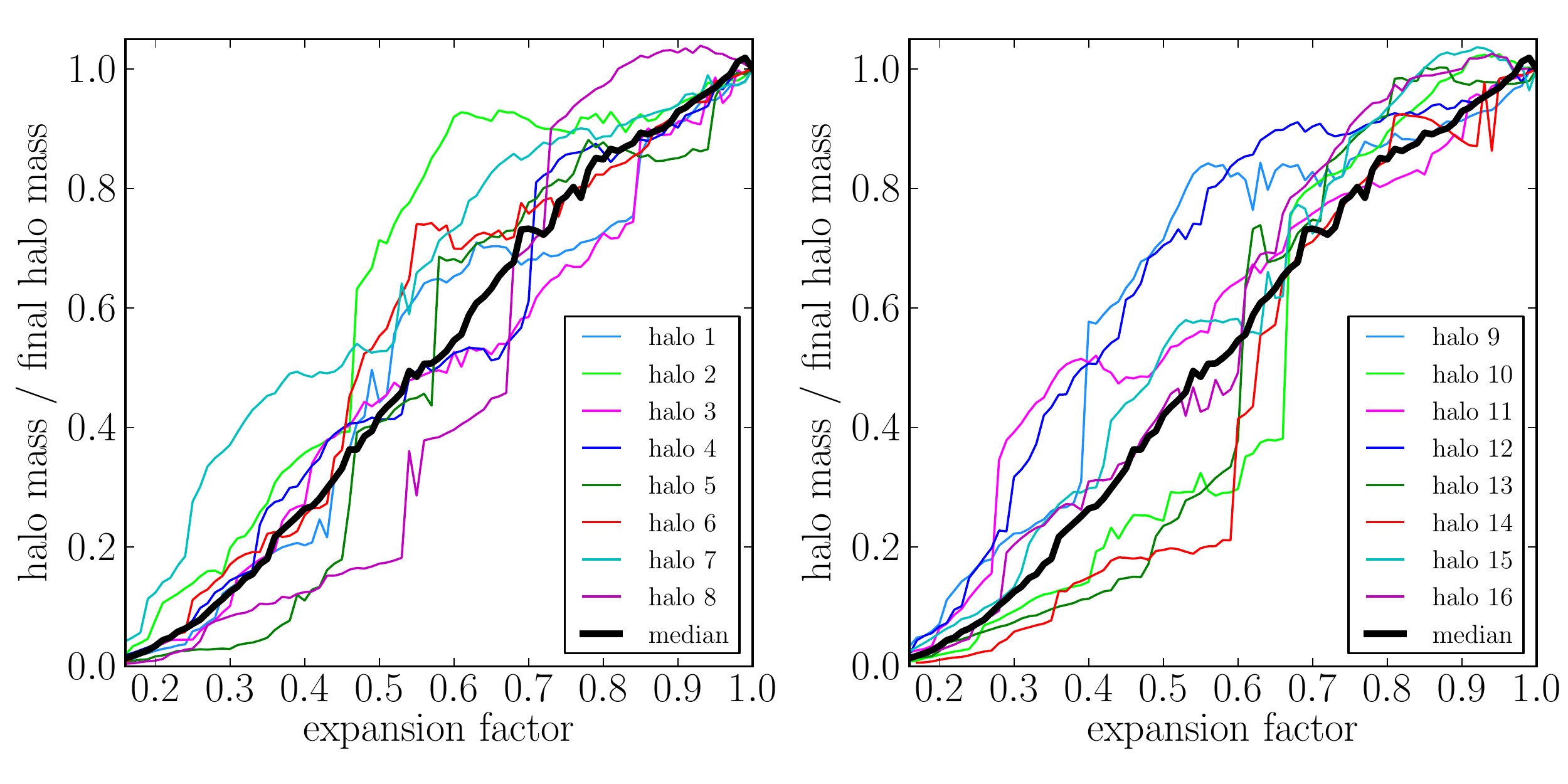}
\caption{Halo mass as function of the expansion factor, in our N body unigrid run. This plot shows the mass accretion history of each of the 16 halos in our sample. The black and thicker curve (labelled median) is the median of a larger sample of 108 halo within the same mass range.}
\label{Time_Evolution_Plot}
\end{figure*}

\begin{figure*}
\centering
\includegraphics[scale=0.7]{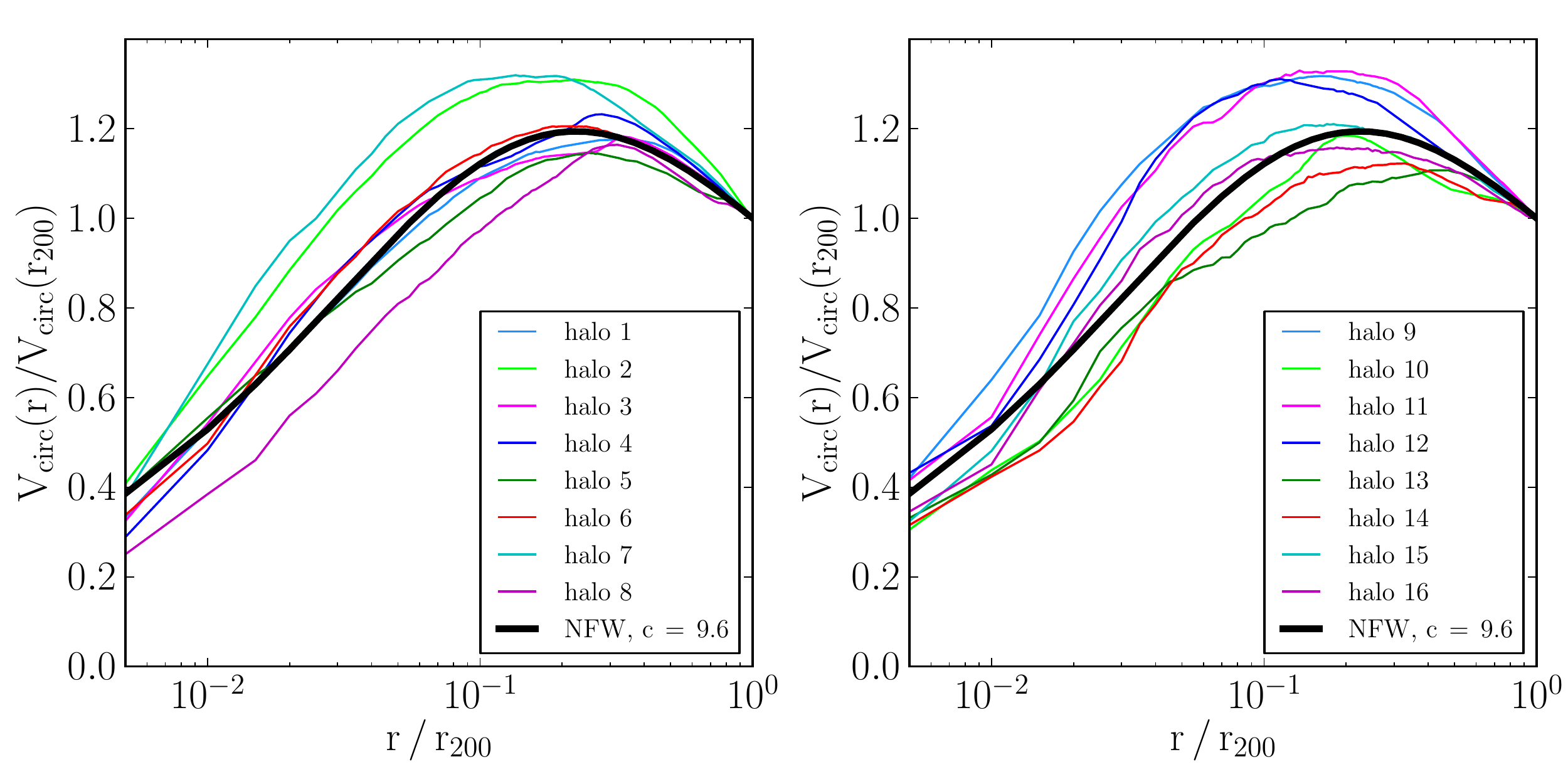}
\caption{Circular velocity profile, in our N body unigrid run. This plot shows the circular velocities of the 16 halos in our sample. The black and thicker curve (labelled analytic) is the NFW model \citep{1996ApJ...462..563N} with a concentration parameter equal to $c=9.6$, typical of our halo mass range \citep{2011ApJ...740..102K}.}
\label{Vcirc_Plot}
\end{figure*}

\begin{figure*}
\centering
\includegraphics[scale=0.7]{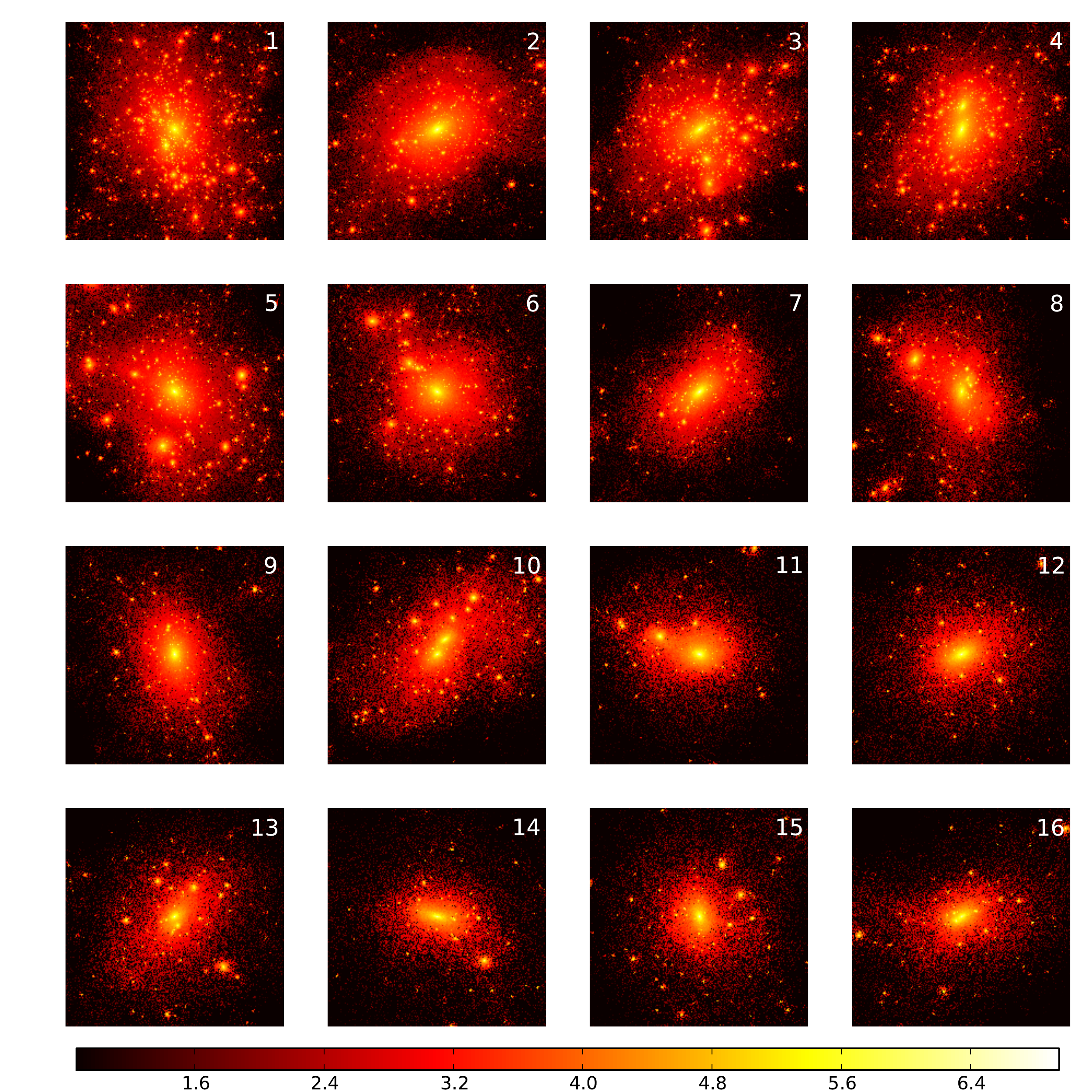}
\caption{Dark matter density maps in our N body zoom-in runs. The colour bar unit is $\mathrm{log}_{10}(\rho_m/\bar{\rho}_m)$, where $\bar{\rho}_m$ is the average matter density in the universe. The side length of each map is $2R_{200}$, with the centre of the maps corresponding to the centre of mass of the halo.}
\label{DM_Density_Maps}
\end{figure*}

\begin{figure*}
\centering
\includegraphics[scale=0.7]{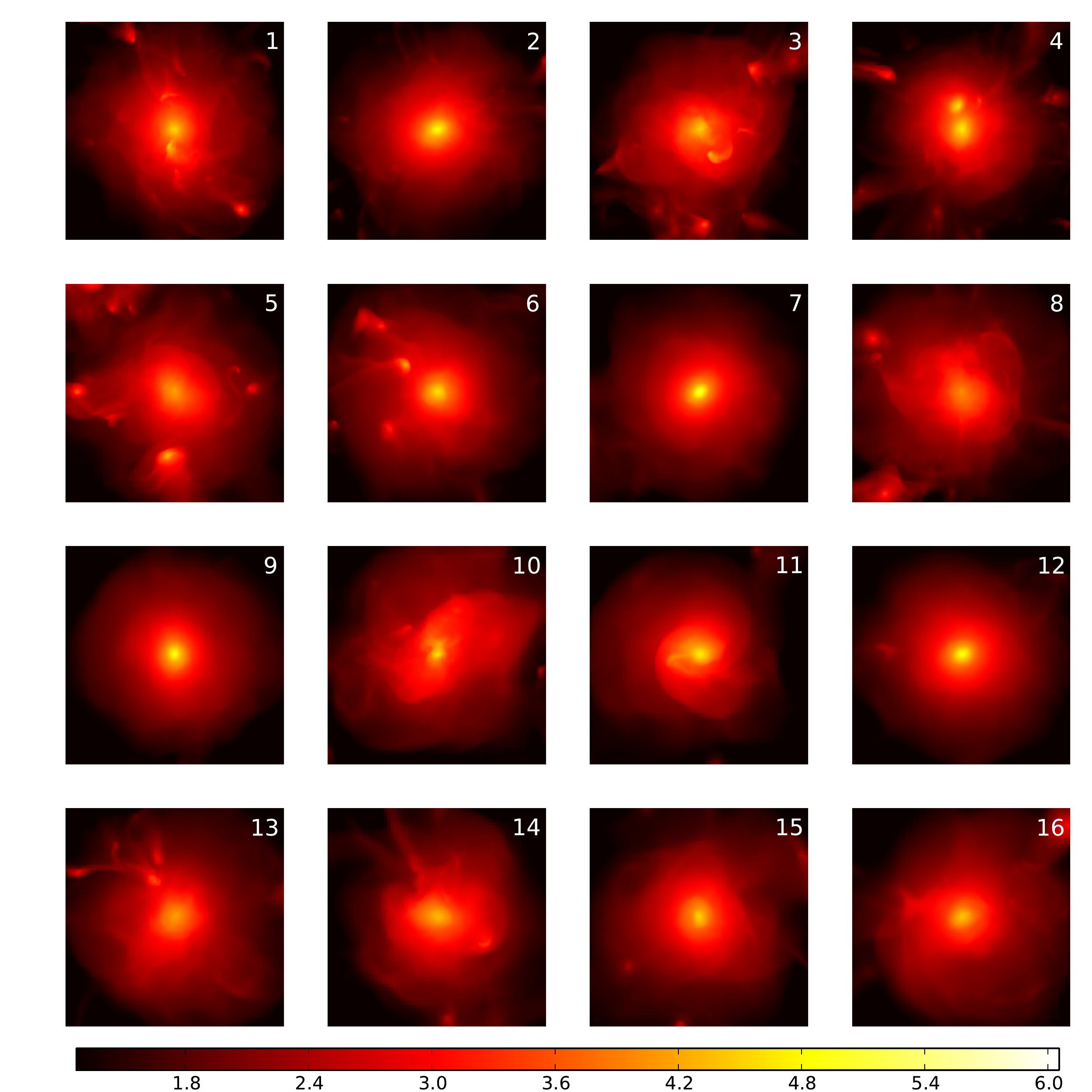}
\caption{Gas density maps in our non-radiative hydrodynamics zoom-in runs. The colour bar unit is $\mathrm{log}_{10}(\rho_{gas}/\bar{\rho}_b)$, where $\bar{\rho}_b$ is the average baryon density in the universe. The side length of each map is $2R_{200}$, with the centre of the maps corresponding to the centre of mass of the halo.}
\label{Gas_Density_Maps}
\end{figure*}

In Figures~\ref{Time_Evolution_Plot} and \ref{Vcirc_Plot}, we plot the mass accretion history and circular velocity profile of our 16 selected halos. One can see from these two plots that halos which have assembled their mass early on, are also systematically more concentrated at $z=0$. 
This confirms the standard scenario of halo formation \citep{1997ApJ...490..493N, 2013ApJ...763...70W, 2013ApJ...767...23W}, in which halos which had their last major merger early, had more time
to absorb the substructure induced from smaller halos falling in on them.  Whereas halos with a late last major merger are still in the process of absorbing subhalos into their core through dynamical processes. 
This interpretation is confirmed by the density maps (Fig.~\ref{DM_Density_Maps} and \ref{Gas_Density_Maps}). The halos with an early formation epoch have a clean spherical shape without large substructure. Some of the halos, classified as average, have a significant number of relatively large subhalos, but they also show a prominent high density core. The four halos in our sample with late formation epoch have less large subhalos than the average case, however they also lack a clearly defined high density core. To quantify the time evolution of the halos with a number, we define the formation redshift $z_{form}$, as the redshift at which a halo has acquired 50\% of its final mass at $z=0$.

\begin{figure*}
\centering
\includegraphics[scale=0.7]{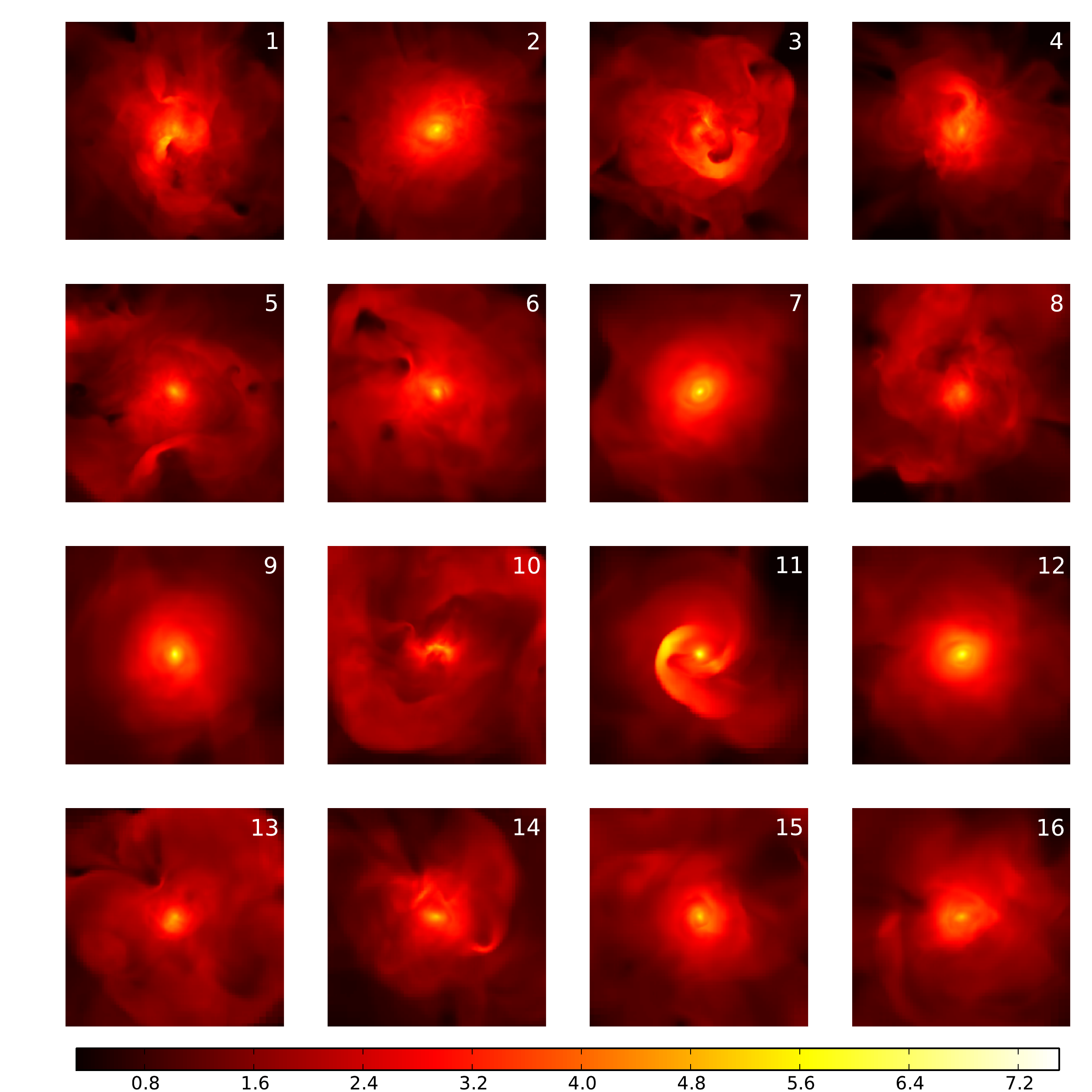}
\caption{Gas temperature maps in our non-radiative hydrodynamics zoom-in runs. The colour bar unit is $T_{gas}/T_{200}$, where $T_{200}$ is computed for each individual halo as $T_{200}=\frac{1}{3}\frac{m_p}{k_B}\frac{G\,M_{200}}{R_{200}}$. The side length of each map is $2R_{200}$, with the centre of the maps corresponding to the centre of mass of the halo.}
\label{Gas_Temperature_Maps}
\end{figure*}

A special case is depicted by halo 10. Although on the gas density map, it looks like a clean case without subhalos, it actually consist of two subhalos of similar mass in the process of merging, clearly visible in the dark  matter projection map. Although it was identified by our halo finder as a relatively isolated halo in our adopted mass range, we did not take into account this exceptional case for our calculations of mean quantities. Nevertheless, we kept it in our sample as a typical example of a statistical outlier and quantified the deviation of its properties from the average of the other 15 halos. 
The exceptional dynamical state of halo 10 can also be seen in the temperature map (Fig.~\ref{Gas_Temperature_Maps}), as a thin shock in between the two subhalos.  For the other halos in our sample, the temperature maps show also typical signatures of strong shocks, but at larger distances from the halo centre. Old halos (like halo 2, 7, 9 and 12) exhibit a very regular temperature structure, with a hotter core and a steady, quasi-spherically symmetric decline towards the external regions. Halo 11, on the other hand, formed early but suffered from a relatively late and massive merger, that can be seen nicely in the temperature map.

\begin{table*}
\centering 
\begin{tabular}{c  c  c  c  c  c  c  c  c}
\\
\hline
Designation	& $M_{200}$ (hydro)  &	$r_{200}$ (hydro) & c (hydro)  &  $r_{200}$ (N-body) & c (N-body)  & $z_{\rm form}$   & assembly	& mass profile \\
		& [$10^{13} M_{\odot} h^{-1}$] &	[Mpc] 		  & 	& [Mpc]	&	& & &		  \\
\hline
1		& 5.59  & 1.36  & 9.31 & 1.37  & 8.48 & 0.96	& average & average	 \\	
2 		& 5.31  & 1.34  & 12.9 & 1.31  & 13.5 & 1.17	& early   & concentrated \\
3 		& 4.27  & 1.24  & 8.91 & 1.25  & 12.2 & 0.69	& average & average	 \\
4 		& 4.04  & 1.22  & 10.7 & 1.2   & 12.3 & 0.79	& average & average	 \\
5		& 3.07  & 1.11  & 8.79 & 1.1   & 8.69 & 0.75	& average & shallow	 \\
6		& 2.11  & 0.982 & 10.6 & 0.972 & 11.5 & 1.13	& average & average	 \\
7		& 1.65  & 0.904 & 16.4 & 0.894 & 16.2 & 1.22	& early   & concentrated \\
8		& 1.56  & 0.888 & 6.85 & 0.862 & 8.58 & 0.49	& late    & shallow	 \\
9		& 1.46  & 0.869 & 14.1 & 0.859 & 14.7 & 1.5	& early   & concentrated \\	
10		& 1.16  & 0.804 & 8.27 & 0.791 & 7.63 & 0.51	& late    & shallow	 \\
11		& 1.01  & 0.768 & 13.9 & 0.729 & 18   & 1.7	& early   & concentrated \\
12		& 0.789 & 0.706 & 14.5 & 0.7   & 16   & 1.56	& early   & concentrated \\
13		& 0.713 & 0.684 & 7.67 & 0.68  & 7.77 & 0.64	& late    & shallow	 \\
14		& 0.616 & 0.651 & 9.57 & 0.622 & 9.85 & 0.59	& late    & shallow	 \\
15		& 0.625 & 0.655 & 10.4 & 0.641 & 11   & 1.04	& average & average	 \\
16		& 0.537 & 0.622 & 9.85 & 0.622 & 10.7 & 0.67	& average & average	 \\
\hline	
\end{tabular}
\caption[Halo Properties]{Properties of our halo sample}
\label{Halo_Table}
\end{table*}


\section{Results}
\label{results}

In this section we present the results of our numerical experiment in the form of profile plots for the relevant thermodynamical and dynamical quantities. Our main points of interest are:
\begin{itemize}
\item how strongly do  our results deviate from the analytical model we have adopted?
\item how strongly do individual halos deviate from the mean?
\end{itemize}
We therefore plot the mean of our 15 halos (excluding halo number 10) and the analytical curve together in the same plots, and quantify the variance of our sample by
\begin{equation}
\sigma=\sqrt{\frac{1}{N}\sum_{i=1}^{N}(x_i-\overline{x})^2}
\end{equation} 
and plot it as a shaded region around the mean curve. Furthermore, we estimate the error in our estimation of the mean by
\begin{equation}
\sigma_{mean}=\frac{\sigma}{\sqrt{N}}
\end{equation} 
This quantity is plotted as the error bars around the mean. The profiles where sampled with 109 radial bins in the range $ 0 < r < 1.5\,r_{200}$ for each halo and for all quantities, unless stated otherwise. The mean and standard deviation are also calculated at the same 109 coordinates. Values of $r$ below the effective resolution of $\Delta x_{min}=L/2^{\ell_{max}}=1.1258\,\mathrm{kpc}$ are discarded. 
\begin{figure*}
\centering
\includegraphics[scale=0.575]{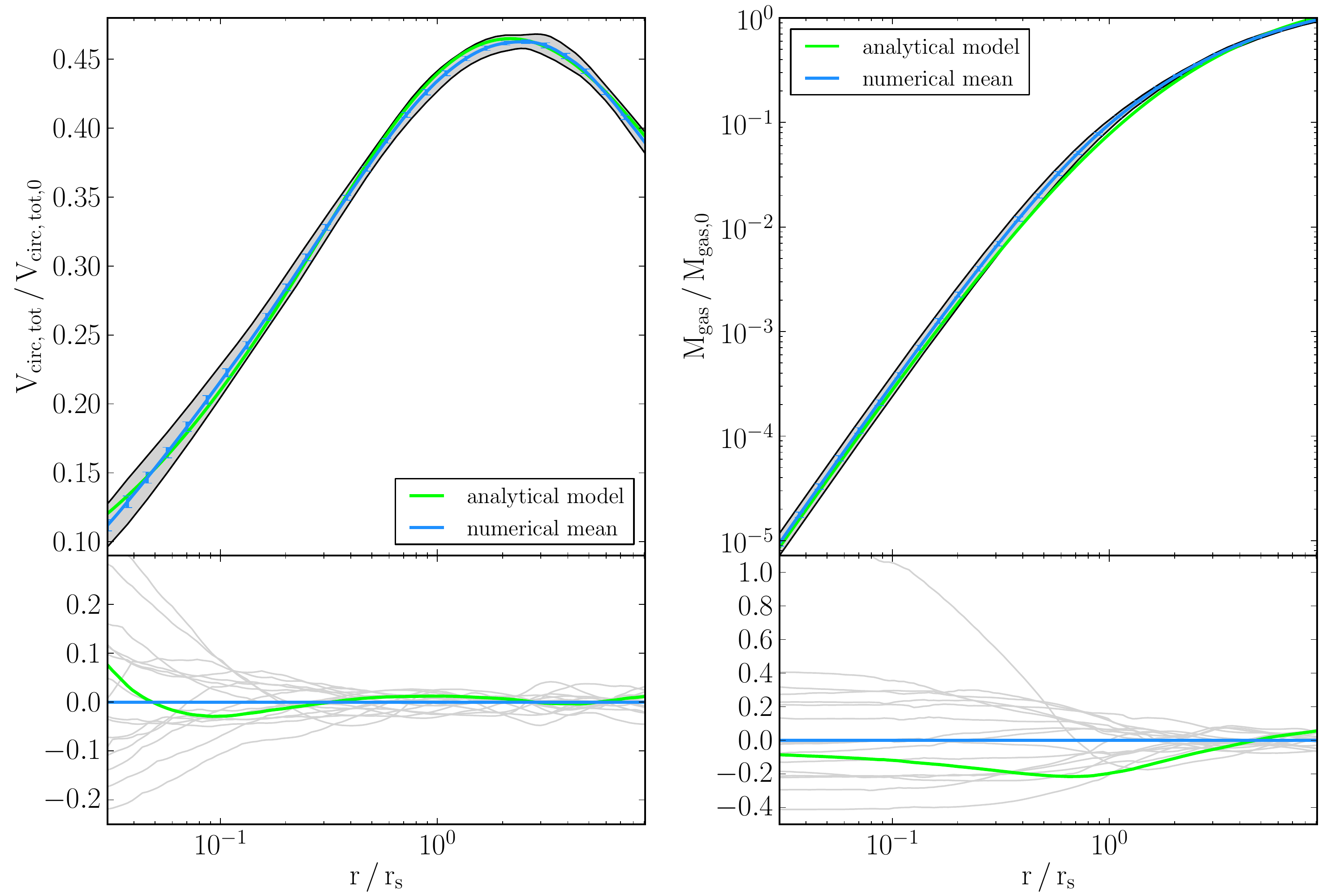}
\caption{Total circular velocity and gas mass profiles: In the upper plot the blue curve with error bars is the numerical mean of the 15 halos, the grey shaded area is the variance of the numerical mean and the blue error bars indicate the error of the mean. The analytic prediction is plotted in green. In the lower panel the deviations from the numerical mean are plotted. Green is again the analytic value and the individual halos' deviations are depicted as thin grey lines.}
\label{Circular_Velocity_Gas_Mass_Plot}
\end{figure*}

\begin{figure*}
\centering
\includegraphics[scale=0.575]{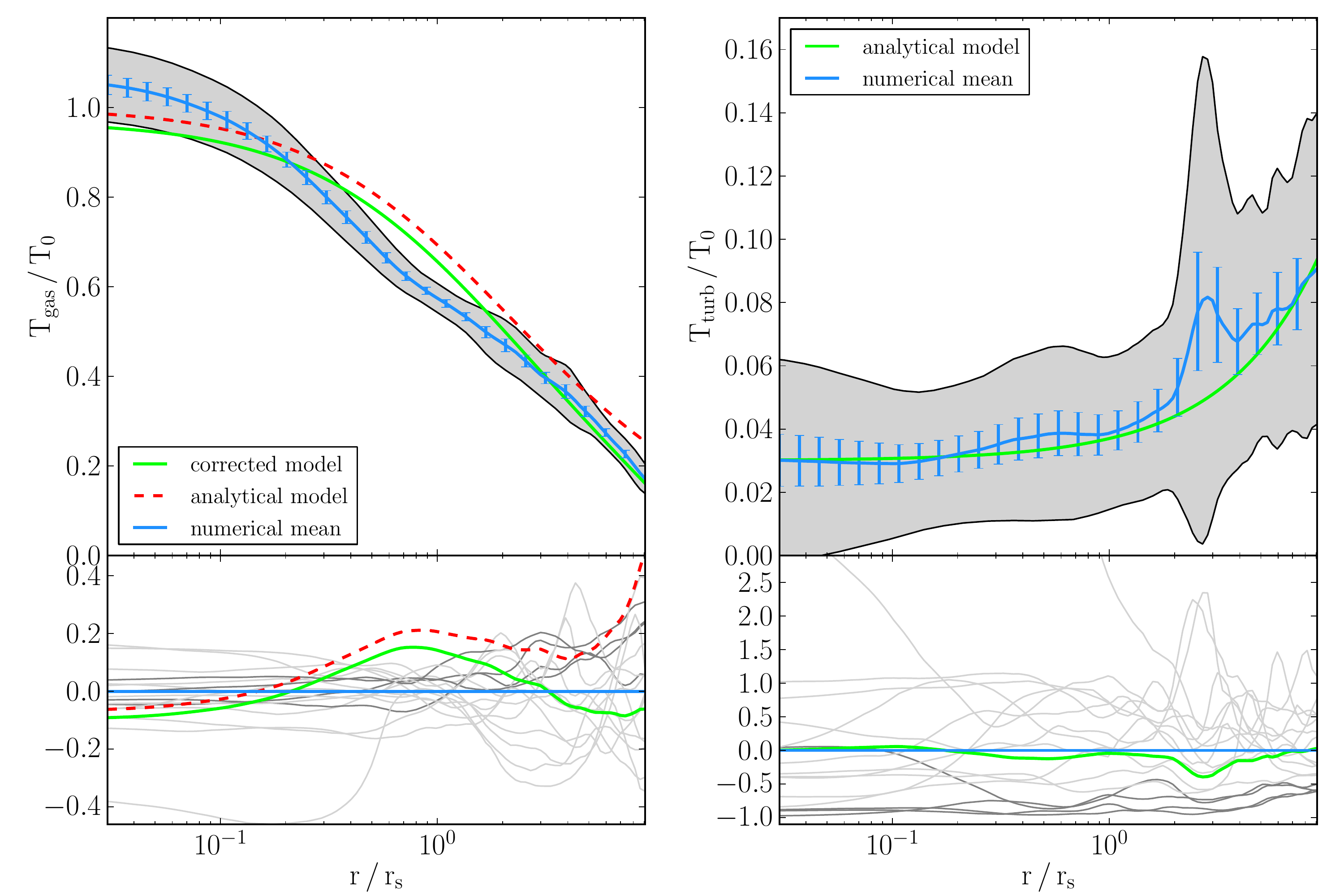}
\caption{Gas temperature and turbulent gas temperature profiles: The green line in the turbulent temperature plot is the fitted analytical model and the green line in the temperature plot is the corresponding corrected analytical model. In addition the uncorrected analytic curve of the gas temperature is plotted in dashed red, in the plot of the gas temperature. The colour code of the individual halo deviations is explained in the text.}
\label{Thermal_Turbulence_Gas_Plot}
\end{figure*}    
\begin{figure*}
\centering
\includegraphics[scale=0.575]{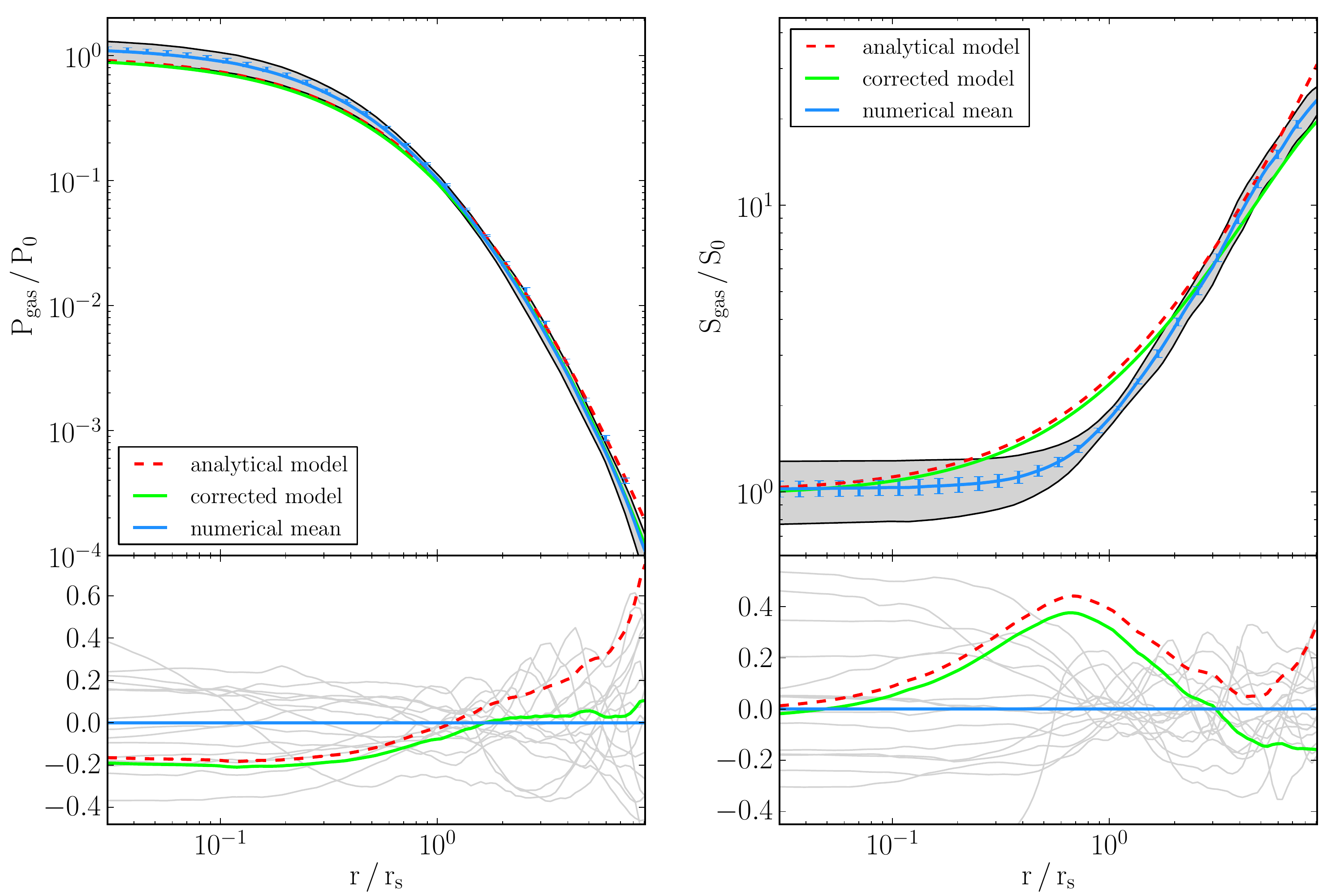}
\caption{Gas pressure and gas entropy profiles: The green line in the pressure plot is the corrected analytical model. In addition the uncorrected analytic curve of the gas temperature is plotted in dashed red. The colour code of the entropy profile is the same.}
\label{Pressure_Entropy_Gas_Plot}
\end{figure*}     
\begin{figure*}
\centering
\includegraphics[scale=0.575]{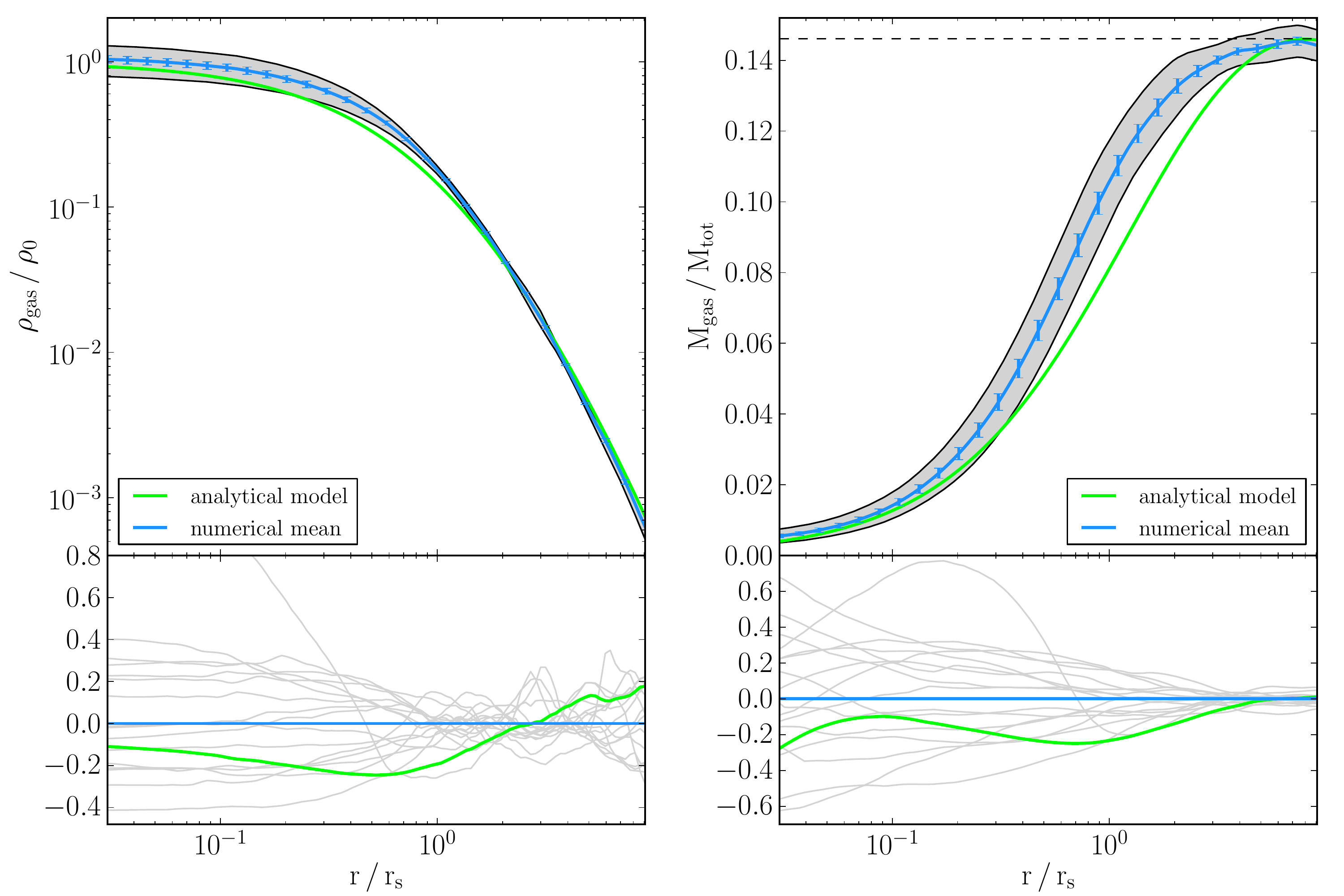}
\caption{Gas density and mass fraction profiles: The colour code is the same as in Figure \ref{Circular_Velocity_Gas_Mass_Plot}. For a better comparison with observational results, we have plotted the unnormalized gas mass fraction profile (right panel). The analytical curve of it is plotted for the case of $c=8$. The dashed line indicates the universal mass ratio of baryons to total matter.}
\label{Gas_Density_Mass_Frac_Plot}
\end{figure*}      

\subsection{Circular Velocity Profile}

To characterise the radial distribution of the total mass found in our halos, we plot the circular velocity normalised to $4 \pi G \rho_s r_s^2$ versus the radius r normalised to $r_s$ in Figure~\ref{Circular_Velocity_Gas_Mass_Plot}. Our mean value is in excellent agreement with the NFW analytic model.
The deviation of the mean from the analytic curve lies below 2\% for $r>0.3\,r_s$. For smaller $r$, it increases up to 7.5\%. 
The deviations of individual halos from the analytical model are below 10\% for $r > 0.1\,r_s$. For $r < 0.1\,r_s$, they increase, but do not exceed 30\%.
The variance in the total circular velocity profile reaches 30\% of the mean at the centre and decreases to less than 4\% in the outer regions.
This result proves the constrained nature of the collisionless dark matter component in cosmological simulations.
We conclude that, in our simulations, the total mass distribution follows the one predicted by the NFW model. 
The only small deviation of individual halos from the mean can be interpreted as a resolution effect on very small scales. 

\subsection{Cumulative Gas Mass Profile} 

The baryonic mass  profile is the other integrated, cumulative quantity, and is in very good agreement with the analytical prediction (Fig.~\ref{Circular_Velocity_Gas_Mass_Plot}).
The deviation of the mean profile from the analytical model lies below 20\% and reaches its largest value at $r_s$. 
The deviations of each individual halos from the mean are all similar, except for one strong outlier, halo 10. 
For $r<0.1\,r_s$, these deviations appear as a constant offset, which for the strongest case corresponds to 40\% above or below the mean profile. 
With increasing radius, individual deviations gradually decrease to 10\% for $ r>r_s$. 
As a consequence, the variance in the gas mass profile decreases monotonically from 50\% of the mean at the centre to less than 1\% in the periphery ($r>2\,r_s$).   

\subsection{Gas Density}
\label{density}

The gas density profile is shown in the left panel of Figure~\ref{Gas_Density_Mass_Frac_Plot}, and exhibits a larger deviation to the mean than the cumulative mass profile. 

The overall agreement with the analytical model is quite good. On closer look, however, it appears clearly that for $r<2\,r_s$, the analytic model underestimates systematically the simulated halo profiles, while for $r >2\,r_s$, the situation is reversed, and the analytical prediction systematically overestimates our numerical results. The deviation of the mean from the analytic profile is strongest at intermediate radii (around $r=0.5\,r_s$) with 25\%, where it appears larger than the standard deviation. This deviation of the analytical profile from our numerical mean is therefore significant around $r_s$ and should be taken seriously.

The deviations of individual halos from the mean have a maximum value of 50\% at the centre (not considering our outlier halo 10), in the form of constant offsets. In the range $r>r_s$ the deviations are smaller (around 20\% to 30\%), mostly in the form of random peaks for each of the 16 halos. This translates into a standard deviation of 35\% in the periphery and 50\% in the centre.

The second feature can be explained by the existence of substructure in the outer parts of the halos. The first one is less obvious. We interpret it as different levels of entropy after the last major mergers, due to different circumstances occurring at halo formation time. This argument was used by \cite{2015arXiv150904289H} to explain the dichotomy between cool core and non-cool core clusters. In our case, low angular momentum mergers would give rise to almost head-on collisions and higher post-shock entropy levels, resulting in a systematically higher temperature and lower density in the core. We will come back to this point in Section~\ref{correlations}.

We find that our numerical average has a typical density core in the centre. In the non-radiative hydrodynamics simulations of the nIFTy comparison project by \cite{2016MNRAS.457.4063S}, 
this feature is shown to be typical for grid-based and modern SPH codes. The same authors report that classical SPH codes give gas density profiles which rise all the way to the centre, 
leading to a disagreement with grid-based codes of one order of magnitude. 

Furthermore, our gas density profile is in agreement with the one from the earlier study of \cite{2003MNRAS.346..731A}, where a modern, entropy conserving version of the SPH code GADGET was used.
They also report a good agreement with the same analytical model, around 20\%. 
However, their gas density profile deviates mostly in the outer parts, whereas our numerical mean deviates mostly in the intermediate range $r \simeq r_s$. 

\subsection{Gas Temperature Profile}
\label{temperature}

We have computed for each halo the temperature profile by averaging the temperature from individual AMR cells within the same 109 spherical bins (with a mass-weighted average).
We have then computed the mean of all 15 profiles (again excluding halo 10), as well as the variance. 
These are shown on Figure~\ref{Thermal_Turbulence_Gas_Plot} and compared to the analytical prediction (see Eq.~\ref{Temp}).
One clearly sees a small but systematic difference between the analytical model (red dashed curve in Fig.~\ref{Thermal_Turbulence_Gas_Plot}) 
and our measurement. As explained in \cite{2010ApJ...725.1452S} and \cite{2012ApJ...758...75B}, we argue this is due to the contribution of the turbulent pressure in the halo, which is missing in the analytical model. 
To quantify the effect of turbulence, we have computed the turbulent energy profile, by averaging in each spherical bin the mass-weighted velocity dispersion as
\begin{equation}
v^2_{\rm turb}(r)=\frac{1}{3}\frac{1}{M}\int \,(v _x^2+v_y^2+v_z^2)\, \rho dV  
\end{equation} 
where the $v_i$ are the components of the velocity of the gas in the frame of the halo centre of mass. We have then converted this velocity dispersion into a turbulent temperature as
\begin{equation}
T_{\rm turb}(r)=\frac{m_h}{k_B} \cdot v^2_{\rm turb}(r)  
\end{equation} 
The average turbulent temperature profile (with error bars) and its standard deviation are both plotted on Figure~\ref{Thermal_Turbulence_Gas_Plot}. 
To first order, the turbulent specific energy appears as constant, with a mean value around $0.06\,T_0$. On closer look, it is in fact slightly rising with radius, 
with a central value around $0.03\,T_0$, reaching in the outer parts $0.10\,T_0$. We propose the following fitting function for the turbulent temperature 
\begin{equation}
T^{\rm fit}_{\rm turb}(r)=\left( 0.03  + 0.007 \frac{r}{r_s} \right) T_0,
\end{equation}
shown as the green solid line in the right panel of Figure~\ref{Thermal_Turbulence_Gas_Plot}.
When one now subtracts from the analytical temperature profile this fit for the turbulent, non-thermal energy 
\begin{equation}
T^{\rm cor}_{\rm gas}(r)=T_{\rm ana}(r)-T^{\rm fit}_{\rm turb}(r),
\end{equation}
one gets much better agreement with our numerical data (see the green solid curve in the left panel of Fig.~\ref{Thermal_Turbulence_Gas_Plot}), 
especially in the outer parts where turbulence is the strongest.

The final agreement between our corrected analytical model and the numerical data is quite good (less than 15\%) but is far from perfect. The maximum disagreement lies around $r_s$ and peaks at 15\%. At this radius, the numerical gas temperature lies significantly below the analytical prediction. It is worth noticing that this radius corresponds also to the maximum discrepancy observed in the gas density profile, where this time the numerical gas density lies significantly above the analytical prediction.

Individual halo temperature profiles deviate from the mean in the centre with an almost constant offset, never exceeding 15\%. This constant offset in the central temperature
is related to the constant offset we have discovered in the central density. This is again due to different merger circumstances at halo formation time (see Section~\ref{density}).
In the outer parts, $r>r_s$ the deviations are characterised by peaks and troughs, which we associate with the presence of both substructure and turbulence. 
At these radii, the deviations from the mean can reach 35\% in the extreme cases.

The variance in the gas temperature profile is also constant in the centre with a value close to 16\% of the mean for $r<0.5\,r_s$ and decreases from there to less than 10\% at $r \simeq r_s$, only to increase again to 35\% around $r \simeq 3\,r_s$. Finally it reduces to 10\% in the outermost range. 

The temperature profiles measured in \cite{2003MNRAS.346..731A} show a much better agreement with the hydrostatic analytical model.
The variance of the temperature profile appears also much more regular and slightly smaller (around 20\%). 
From a first naive look, a possible interpretation for the smoother profiles in these earlier SPH simulations could be coming from the selection strategy in \cite{2003MNRAS.346..731A}, where major mergers were first identified and then removed from the sample. 
In our work only halo 10 classifies as a major merger in this sense. And since we excluded it from the computation of the numerical average (as mentioned above), the difference in the temperature variance, between the two works, cannot arise from the removing of major mergers.
For the better agreement with the uncorrected analytical profile, we argue that the SPH method used in \cite{2003MNRAS.346..731A} is known to dissipate subsonic turbulence too  efficiently.
Indeed, as demonstrated nicely in \cite{2012MNRAS.423.2558B},standard SPH techniques underestimate the turbulent energy by a factor between 2 and 10, depending on the scales considered. This reduces artificially the non-thermal pressure support,
especially in the outer parts where turbulence is the strongest, and as a consequence provides a better but spurious match to the strict hydrostatic model. Interestingly, \cite{2002ApJ...579..571L}
have also found a universal temperature profile (using mock X-ray observations) using non-radiative AMR simulations similar to ours. \cite{2003MNRAS.346..731A} have directly compared their SPH simulations
to the AMR simulations of  \cite{2002ApJ...579..571L} and have found a systematic 10\% positive difference between SPH and AMR results, explained by the stronger turbulence support in the AMR case.

\subsection{Turbulence and substructure}

As already presented in the previous section, the level of turbulent energy in our simulated halos is relatively low and uniform, between 3\% and 10\% of the thermal energy, estimated here as the central temperature $T_0$. This is agreement with \cite{2009ApJ...705.1129L} and \cite{2016arXiv160804388M}, who report very similar values. Note that for the latter work, the authors used a model with a constant turbulent specific energy, while in our case, we see a clear, although not very strong, increase of the turbulence specific energy with radius, confirming the analytical theory presented in \cite{2014MNRAS.442..521S} and \cite{2015MNRAS.448.1020S}. 
From individual halo turbulent profiles (see Fig.~\ref{Thermal_Turbulence_Gas_Plot}), one can see pronounced peaks in the profiles for $r>r_s$.
This is a direct consequence of substructure in the outer regions of the halos. 
This translates in a very large variance for the turbulence, usually up to 100\%, but even larger in some cases. 

In our catalog, we have a sub-sample of 4 halos with very low levels of turbulence: halos 2, 7, 9 and 12, which all show very smooth gas density maps (Fig.~\ref{Gas_Density_Maps}) and were classified as early formation epoch. 
We have highlighted this subset (halo 2, 7, 9 and 12) with a darker grey line in Figure~\ref{Thermal_Turbulence_Gas_Plot}, to show that these four halos have a significant and almost constant deficit of turbulence compared to the mean. 
This comparison directly indicates a strong correlations between turbulence and substructure on one hand, and concentration and formation epoch, on the other hand. The latter is because halos which have had their last major merger early have had more time to see the turbulence decay. As a result, these halos are well virialised, with low levels of turbulence and have temperature profiles closer to the uncorrected analytical model. 
To support further these conclusions, we have investigated possible correlations between the thermodynamical properties, the concentration parameter $c$ and the formation redshift $z_{\rm form}$ in Section~\ref{correlations}.

The complementary viewpoint our outlier halo 10, which consist of two subhalos of similar mass in the merging process at $z=0$. 
In this case, the gas temperature in the centre shows a deficit with respect to the mean as high as 50\%. 
The turbulent energy, on the other hand, overshoots the mean value by a factor of 10, meaning that in the case of this halo, the turbulent energy in the centre is almost 30\% of $T_0$, accounting for almost all the missing thermal energy, which is needed to balance the HSE situation.  
This kinetic energy is stored in the velocity of the relative motion of the two merging sub-halos, waiting to be dissipated into heat when their orbital separation will shrink to zero.
This shows, that when the kinetic energy of the turbulence is taken into account, even halos which are far from being virialised (like halo 10), can still be compared to a model of hydrostatic equilibrium.

\subsection{Pressure}

Since the gas pressure is proportional to the product of gas temperature and gas density, it could be considered in a sense as a redundant quantity. 
We nevertheless plot it in Figure~\ref{Pressure_Entropy_Gas_Plot} to emphasise three distinct features.

First, while it shows quantitatively the same behaviour as the gas density, the variance being significantly smaller in the central region. 
Second, the agreement of our mean profile with the corrected analytical model is quite good, especially in the outer parts $r>r_s$ (less than 10\%), and still good in the inner parts $r<r_s$ with less than 20\%.
We do not see a significant disagreement particularly around $r\simeq r_s$, like for the density and the temperature profiles. 
Third, the peaks and the troughs in the outer regions are more pronounced than for the density and the temperature, as they reach 60\% for the pressure, while for the density (resp. the temperature) they reach only 20\% (resp. 35\%).

We interpret the first two features as coming from the close connection between the gas pressure and the total mass distribution through the hydrostatic equilibrium equation. The small deviations in the pressure profile reflect the small deviations in the circular velocity profile of the total mass, just as the good agreement with the corrected analytical model for the pressure reflects the good agreement with the analytical model of the total mass circular velocity. 
The third feature is due to shocks associated to sub-structure collapsing through the main halo and converting their kinetic energy into heat. 

The difference between the corrected and uncorrected analytical analytical model is barely visible for $r<r_s$ (less then 5\%). But it increases significantly in the outer regions, where it reaches 60\%. 
The obvious interpretation is the existence of an additional pressure support in the form of turbulent pressure in the outer parts, which is not taken into account by the original analytical model, but which is captured correctly by the corrected version (see Section~\ref{temperature}).

\subsection{Gas Entropy Profile}

The gas entropy profile, highly relevant for X-ray observations, is defined as
\begin{equation} 
S_{\rm gas}(r)=\frac{T_{\rm gas}(r)}{\rho_{\rm gas}(r)^{2/3}}
\end{equation}
and is plotted in Figure~\ref{Pressure_Entropy_Gas_Plot}.
We find two different regimes: a constant entropy core for $r < 0.5\,r_s$ and a steep entropy increase for $r>0.5\,r_s$. 
This increase of the entropy in the outer parts is predicted by our analytical model, 
but it fails at reproducing the sharp transition towards a flat core that we observe in our simulations.

Nevertheless, the agreement between the model and our numerical mean is quite good, it shows again a significant (40\%) deviation around $r_s$. 
The deviation of each individual halo from the mean profile follows from the structure of the density and temperature profiles: a constant offset in the centre (with up to 50\% deviation for some halos) and several low amplitude peaks and troughs at large radii ($r>r_s$) due to substructures.

The constant entropy core in the centre and the steep increase in the outer parts are consistent with our finding that the relation between the density and the temperature revealed in Figure~\ref{gamma} seems to exhibit two regimes: 
\begin{enumerate}
\item the core regime, at high density, for which the entropy is nearly uniform, as a consequence of the strong mixing following substructure mergers, 
and resulting in a polytropic law of $\Gamma \simeq 5/3$, 
\item the halo regime, at low density, for which the entropy is rising with increasing radius,
as a result of the evolution of the accretion shock leaving behind a stratified, convectively stable atmosphere with $\Gamma \simeq 1.2$. 
\end{enumerate}
This bi-modal, core-halo evolution is at odd with the main assumption of the polytropic analytical model, namely a unique value for $\Gamma=1.19$.
This explains why we see significant deviation between our numerical mean and the analytical model around $r \simeq r_s$.

In \cite{2016MNRAS.457.4063S}, the constant entropy core is reported for all grid-based and modern SPH codes. 
Classical SPH implementations, on the other hand, show a continued steep decrease towards the centre. 
This dichotomy has been discovered first by \cite{1999ApJ...525..554F}, but with the caveat that the resolution was quite limited for most of the participating codes.
In the outer parts, however, all codes seem to agree on the steep increase of entropy.
The grid-based simulation analysis of \cite{2007ApJ...668....1N} confirmed the presence of a flat entropy core in non-radiative simulations.

In the work of \cite{2003MNRAS.346..731A}, the agreement between the measured entropy profiles and the analytical model is better than in our case, around 30\%, 
with the analytical model lying below the numerical profiles.
We believe that these differences are due to different level of entropy mixing in the central region, leading to a less pronounced constant entropy core in the \cite{2003MNRAS.346..731A} results.

\subsection{Cumulative Gas Mass Fraction}

Another important observable for X-ray astronomy is the cumulative gas fraction defined as
\begin{equation}
f_{\rm gas}(< r)=\frac{M_{\rm gas}(< r)}{M_{\rm tot}(< r)}
\end{equation} 
Within our analytical framework, this quantity is not strictly self-similar, and varies very weakly as a function of the concentration parameter $c$.
We plot the gas fraction as a function of the scaled radius in Figure~\ref{Gas_Density_Mass_Frac_Plot}, and compare it to our analytical solution for the specific case $c=8$. Other models with values of c between 4 and 20 are very close (within 2\%) to this reference curve. 

The agreement between the mean profile and the analytical model is quite good, but we see again a clear difference, larger than the standard deviation, in the range $0.5\,r_s<r<2\,r_s$. 
Note that we recover exactly the universal baryon fraction (shown as the horizontal dashed line) already at $r \simeq 3\,r_s$. 
The analytical model satisfies the same constraint by construction (see Section~\ref{analytical_Model}).
The variance in the numerical prediction is quite small, less that 20\%, except again for the outlier halo 10.

When comparing again to \cite{2003MNRAS.346..731A}, we note that their SPH results barely reach 90\% of the universal baryon fraction at $r \simeq 10\,r_s$. 
This is a significant difference with our results.
\cite{2016MNRAS.457.4063S} have reported the same discrepancy between different codes, with SPH codes showing a systematic deficit of baryons at large radii,
while grid-based codes reaching the universal baryon fraction already at relatively small radii, and in most cases even overshooting it. One could speculate that because dark matter is
collisionless, dark matter particle are splashing back to larger radii after their first pericentre passage, and because gas is collisional, it gets shocked and remains trapped at smaller radii.
Both facts combined, this could lead to a deficit of dark matter and an excess of baryon in the halo outer regions, as we observe in our non-radiative simulations and in \cite{2016MNRAS.457.4063S} for grid-based codes. 
 
\subsection{Summary}

We have quantified the dispersion of our profiles with respect to the average profile by measuring the variance with typical values of 10-20\% of the numerical mean. 
In extreme cases, it can reach up to 35-50\%. 
Individual halos profiles deviate up to 20-40\% from the mean, and are mostly in the form of constant offsets in the centre, and in the form of peaks and throughs in the outer regions.   
 
We have also estimated how well the numerical results reproduce the analytical profiles predicted by the model introduced in Section~\ref{analytical_Model}. 
We find an overall good agreement for all quantities. 
In the case of the gas temperature, however, our numerical results significantly underestimated the analytical prediction. 
We argued that we have to include to the pressure support a significant contribution of the turbulence, especially in the outer regions. 
We have fitted the turbulent specific energy with a simple linear function of the radius, and subtracted it from the analytical temperature profile. 
After this correction, the deviations of the analytical model from the numerical mean remain smaller than 20\%.  

We have confirmed the results of \cite{2003MNRAS.346..731A}, namely that the analytical hydrostatic and polytropic gas profiles resulting from an NFW total mass distribution 
(Equations~\ref{Temp} and \ref{Gas_dens}) are good estimates for the actual numerical profiles. 
Note that we have observed a very good agreement between the total mass profile (gas and dark matter combined) and the NFW model. 
We would like to point out, however, that \cite{2003MNRAS.346..731A} fitted the NFW model to the dark matter mass distribution, ignoring the baryons. 
This could partly explain why, in their case, they seem to find an excellent agreement between the numerical temperature profile and the uncorrected analytical model, without the need for invoking turbulence.
We have checked this issue by extracting the NFW parameters $r_s$ and $\rho_s$ from the circular velocity plot of the dark matter mass only, as in the previous work by \cite{2003MNRAS.346..731A}. 
Our assumption was partly confirmed, since we found a better agreement between the uncorrected analytical and numerical curves in the intermediate range $0.5\,r_s < r < 5\,r_s$. 
Above and below this interval however, the differences between the two curves became even larger.  
The other possibility is that their SPH simulations are underestimating by a factor of 2 (or more) the level of residual turbulent energy. 
Note that our mass resolution is higher by a factor of 20 than was achievable more than 10 years ago. 
The size of their sample is similar to ours, with 15 halos, 
but they are distributed over a wider mass range and contain also galaxy cluster sized objects.

We also noticed that \cite{2003MNRAS.346..731A} measured a smaller variance for the profiles than we did. 
A possible underestimation of the turbulence, could explain this discrepancy.
Our results agree also very well with the Figure~1 of \cite{2007ApJ...668....1N} obtained with a sample of 16 galaxy clusters simulated with the Eulerian code ART.
For the gas density and temperature profiles, we have reproduced the behaviour at small radii reported in \cite{2016MNRAS.457.4063S}, for grid based codes and modern SPH codes,
namely a core of constant entropy in the centre, in contrast to this classical SPH codes with an entropy profile decreasing all the way to the centre.

\section{Correlations Between Halos Structural Parameters}
\label{correlations}

\begin{figure*}
\centering
\includegraphics[scale=0.75]{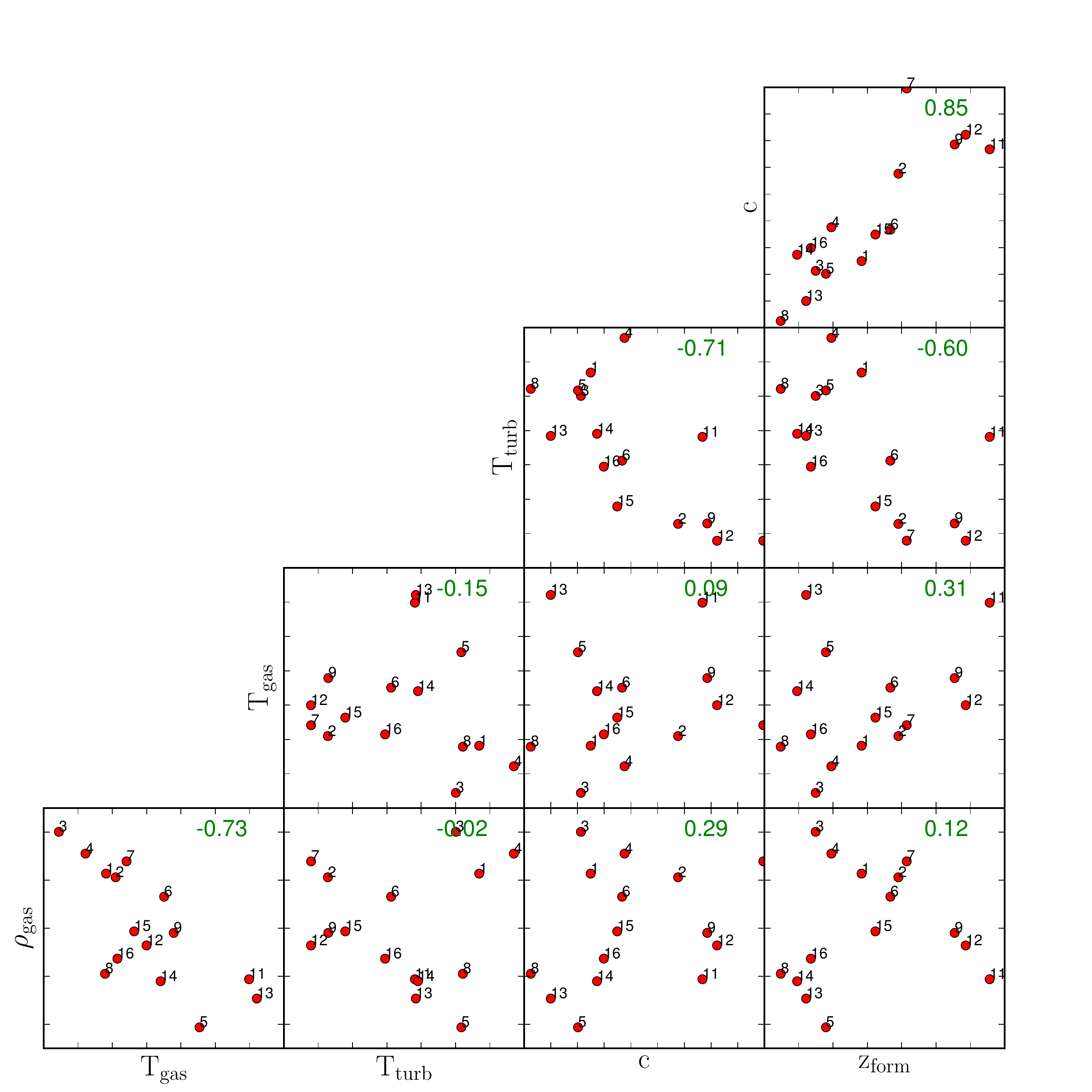}
\caption[]{This plot shows the correlation between the three structural parameters of the gas (central density, central temperature and average turbulent temperature) and the two structural parameters of the halo (formation redshift and concentration). The number in the upper right of each panel is the respective Pearson correlation coefficient, computed without halo 10.}
\label{delta}
\end{figure*} 

In the previous section, we have compared our sample of 16 halos to an hydrostatic analytical profile. For a given halo mass, usually defined by $M_{200}$, one needs to introduce an important structural parameter, namely the concentration parameter $c$. The statistic of this parameters has been well studied using N body simulations \citep{2001MNRAS.321..559B}, and can be considered as an independent random variable. Once $M_{200}$ and $c$ have been chosen, we can deduce the corresponding values for $r_s$ and $\rho_s$, and the hydrostatic equations give us immediately $T_0$ and $\rho_0$ (see Section~\ref{analytical_Model}). 

In order to improve the quality of the fit for a given halo, we now introduce 2 new structural parameters $\rho_{\rm gas}$ and $T_{\rm gas}$, which denotes the central gas density and the central gas temperature. 
We have seen in Section~\ref{density} and Section~\ref{temperature} that each individual halo profile was offset with respect to the analytical prediction $\rho_0$ and $T_0$ by a fixed amount. We interpreted this constant offset in the centre as different entropy levels reached at halo formation time. We now consider these 2 new parameters $\rho_{\rm gas}$ and $T_{\rm gas}$ as two possible independent random variables, and will study now their correlation properties.

In the previous section, we have used the turbulent energy to correct the analytical gas temperature, in order to account for non-thermal pressure support,
and we have identified a correlation between the amount of turbulent energy in each halo and its formation epoch. 
The level of turbulence in the halo is therefore another new and important structural parameter. We define it as 
\begin{equation}
\frac{k_{\rm B}T_{\rm turb}}{m_{\rm H}}=\frac{1}{M(<r_{\rm max})} \, \int_0^{r_{\rm max}} v^2_{turb}(r) \rho(r) 4\pi r^2 dr
\end{equation}
where $r_{\rm max}=10.8\,r_s$ is used as upper bound of the integral because 10.8 is the average $c$ value. For comparison, we have also calculated the integral by using the $r_{200}=c \cdot r_s$ value of each individual halo as upper limit. This had only an insignificant influence on the result. 

We now show the correlation of the various pairs of the following 5 possibly independent random variables $\left(\rho_{\rm gas}, T_{\rm gas}, T_{\rm turb}, c, z_{\rm form} \right)$ in Figure~\ref{delta}.
To quantify the correlations between two random variables, we calculate the Pearson correlation coefficient $C$ and show it in the corresponding panel.

The correlation between $c$ and $z_{\rm form}$ is very high, with a Pearson coefficient of 0.85. This well know properties \citep[see for example][]{2001MNRAS.321..559B} reveals that concentrated halos have formed at an earlier epoch. As we have already anticipated, we also have strong anti-correlations between $T_{\rm turb}$ and $c$ with a Pearson coefficient of $-0.7$ and, similarly between $T_{\rm turb}$ and $z_{\rm form}$ with a correlation coefficient of $-0.6$. 
For the particular cases of halos 2, 7, 9 and 12 (labelled with numbers in Fig.~\ref{delta}), one can see that they form a subset of halos that formed particularly early, with a rather high concentration and a rather low level of turbulence. The opposite is true for halo 8, which formed late, has a low concentration and a large amount of turbulence. 

While there is no correlation between the central gas parameters $\rho_{gas}$ and $T_{gas}$ and the halo structural parameters $c$ and $z_{\rm form}$, the central gas quantities themselves are strongly anti-correlated with a Pearson coefficient of $-0.73$. We interpret this anti-correlation as the consequence of different merger scenarios at halo formation time, with almost head-on collisions leading to higher entropy levels (with a lower density and a higher temperature) and with higher angular momentum collisions leading to lower entropy levels (with a higher density and a lower temperature). In our sample, halo 10 is a prototypical example of such head-on collisions, while halo 11 shows a nice case of a merger with a high angular momentum clump. This interpretation was proposed first by \cite{2015arXiv150904289H} as a possible origin for the cool core/non-cool core dichotomy observed in X-ray clusters.

In conclusion, once we know $M_{200}$, we can draw the concentration parameter $c$ from a log-normal statistic \citep{2001MNRAS.321..559B}. 
We can immediately deduce the expected level of turbulence using the observed correlation between $T_{\rm turb}$ and $c$ in Figure~\ref{delta}. 
We then draw another random variable for $\rho_{\rm gas}$, with a 40\% variance around $\rho_0$, and deduce immediately the central gas temperature 
using the observed correlation between $T_{\rm gas}$ and $\rho_{\rm gas}$ in Figure~\ref{delta}. This strategy could be used to generate a mock catalogue
with very accurate and realistic non-radiative gas properties.  

\section{Effect of Numerical Parameters}
\label{effect_of_numerical_parameters}

We have presented so far the radial profiles of various quantities, comparing the mean value of a sample of 15 halos to a reference analytical profile.
The variance from halo to halo gives the upper envelope of the required accuracy for the analytical model.
We have found that the analytical model deviates significantly (more than the measured variance) from our numerical mean around the scaled radius $r_s$, 
leading us to the conclusion that the single polytropic model is probably too naive,
and does not reflect the bi-modal, core-halo structure of our simulated halos.  

We now want to test the robustness of these conclusions against possible numerical errors. For this, we selected one halo in our sample, halo 2, and re-ran a series of zoom-in simulations, 
varying the following numerical parameters: mass and spatial resolution, type of initial conditions, ingredients of the hydrodynamics solver. 
We then compare the resulting profiles with the fiducial run, using the same quantities discussed in Section~\ref{results}.  
To assess if numerical errors are significant, we again use the variance of the profiles of the 15 halos.

\subsection{Effect of Resolution}

To ensure that our simulation are numerically converged, we ran two additional zoom-in simulations for halo 2, 
where only the maximum level of the initial condition was reduced from our fiducial value $\ell_{\rm ini}=11$ (high resolution), 
down to $\ell_{\rm ini}=10$ (medium resolution) and $\ell_{\rm ini}=9$ (low resolution). 
This translates into mass resolutions of 
$m_{\rm DM}=8.2 \times 10^6 \,M_{\odot} h^{-1}$,  $m_{\rm gas}=1.6 \times 10^6 \,M_{\odot} h^{-1}$ (high resolution)
$m_{\rm DM}=6.6 \times 10^7 \,M_{\odot} h^{-1}$,  $m_{\rm gas}=1.2 \times 10^7 \,M_{\odot} h^{-1}$ (medium resolution) and 
$m_{\rm DM}=5.3 \times 10^8 \,M_{\odot} h^{-1}$ , $m_{\rm gas}=1.0 \times 10^8 \,M_{\odot} h^{-1}$ (low resolution).

\begin{figure*}
\centering
\includegraphics[scale=0.7]{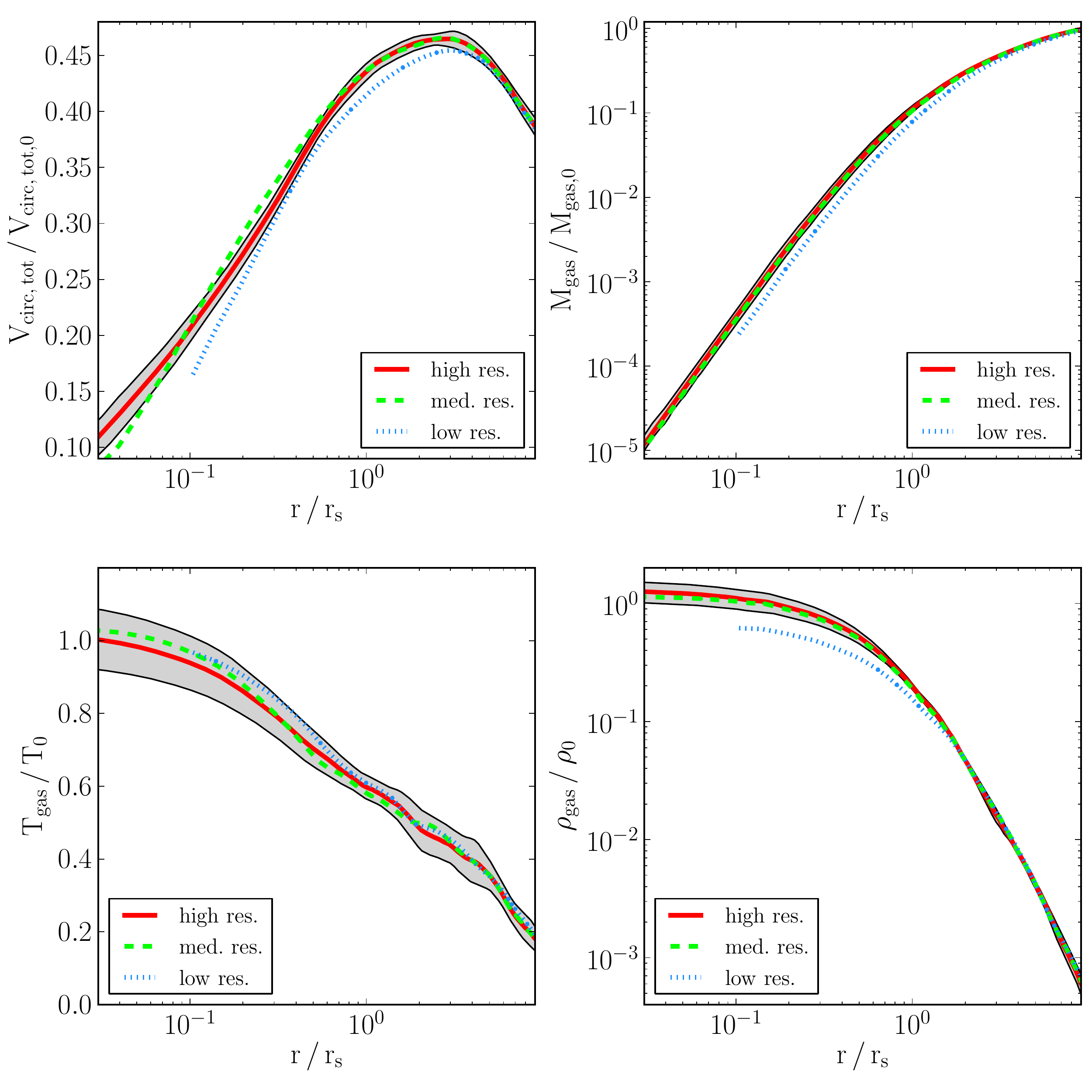}
\caption{The total circular velocity, gas mass, gas temperature and gas density profiles of halo 2 for the three different resolutions under consideration. 
For comparison the variance of the numerical mean of the 15 halos, as it is defined in section 5, is also plotted as grey shaded area.}
\label{Resolution_Hydro}
\end{figure*}

The radial profiles for the runs with non-radiative hydrodynamics are shown in Figure~\ref{Resolution_Hydro}. 
The high resolution profiles are considered here as the reference profiles, and we used the variance over the 15 halos to estimate the required level of accuracy to test for convergence.
One can see that the medium and high resolution runs are both within the shaded area for all quantities, meaning that the measurements are converged within the target accuracy.
Since halo 2 has a mass of  $M_{200}=5.3 \times 10^{13} \,M_{\odot} h^{-1}$, this means we need at least 1 million particles within $R_{200}$ to have fully converged profiles over the radius range 0.1 to 10 in units of $r_s$. In the low mass range of our sample, this requirement is only reached at high resolution, hence validating the adopted resolution for the entire sample.

\subsection{Effect of the Initial conditions}

Our initial conditions were generated using the MUSIC code  \citep{2011MNRAS.415.2101H}. 
Several options are offered to the users of MUSIC to generate the initial particle positions and velocities, as well as the initial gas density and velocity fields.
In this paper, we considered the same transfer function for the combined dark matter and baryonic fluid. We do not explore the possibility to use different transfer functions
for the two fluids, as it is likely to have a small effect on the large scales we consider in this paper. 

We have still the option to compute the particle positions using either the Zel'dovich approximation,
also referred to as First Order Lagrangian Perturbation Theory (1LPT), or the Second Order Lagrangian Perturbation Theory (2LPT). It has been argued \citep{2013MNRAS.431.1866R} that the latter is required for relatively late starting redshift ($z_{\rm ini}\simeq 50$), while 1LPT is enough for early starting redshift ($z_{\rm ini} \ge 100$). Note that for zoom-in simulation using AMR, it is particularly important to start as late as possible,
to make sure that the truncation errors in the potential calculation remain smaller than the initial physical perturbations, justifying the use of 2LPT in this context.

For the gas, we have also two options offered by MUSIC. Either we use the Gaussian random density fluctuations linearly extrapolated to the starting redshift, using in a sense First Order Eulerian Perturbation Theory, or we use the density field corresponding to the adopted Lagrangian Perturbation Theory for dark matter. The second option is referred to a Local Lagrangian Approximation (LLA), and ensures that
the gas density fluctuations are consistent with the slightly non-Gaussian dark matter density field. For more details and references, we point the interested reader to \cite{2011MNRAS.415.2101H}.
 
The various options, 1LPT or 2LPT for dark matter, and with or without LLA for baryons, result in four different combinations summarised in Table~\ref{ICS_Table}. 
For our group catalogue, we use as fiducial choice 2LPT without LLA. We justify this choice in this section, showing that our results are not sensitive to the details in the initial conditions,
given the target accuracy set by the relatively variance in the profiles. For further studies requiring a better accuracy, we argue that the best combination would be however to choose 2LPT with LLA.
 
\begin{table}
\centering 
\begin{tabular}{| c  | c  | c |}
\hline
2LPT 	& LLA &		Run name \\
\hline
yes		& no &		2LPT, no LLA \\
yes 		& yes &		2LPT, with LLA\\
no 		& no &		1LPT, no LLA	\\
no 		& yes &		1LPT, with LLA\\
\hline
\end{tabular}
\caption[MUSIC numerical parameters]{MUSIC numerical parameters}
\label{ICS_Table}
\end{table}

\begin{figure*}
\centering
\includegraphics[scale=0.7]{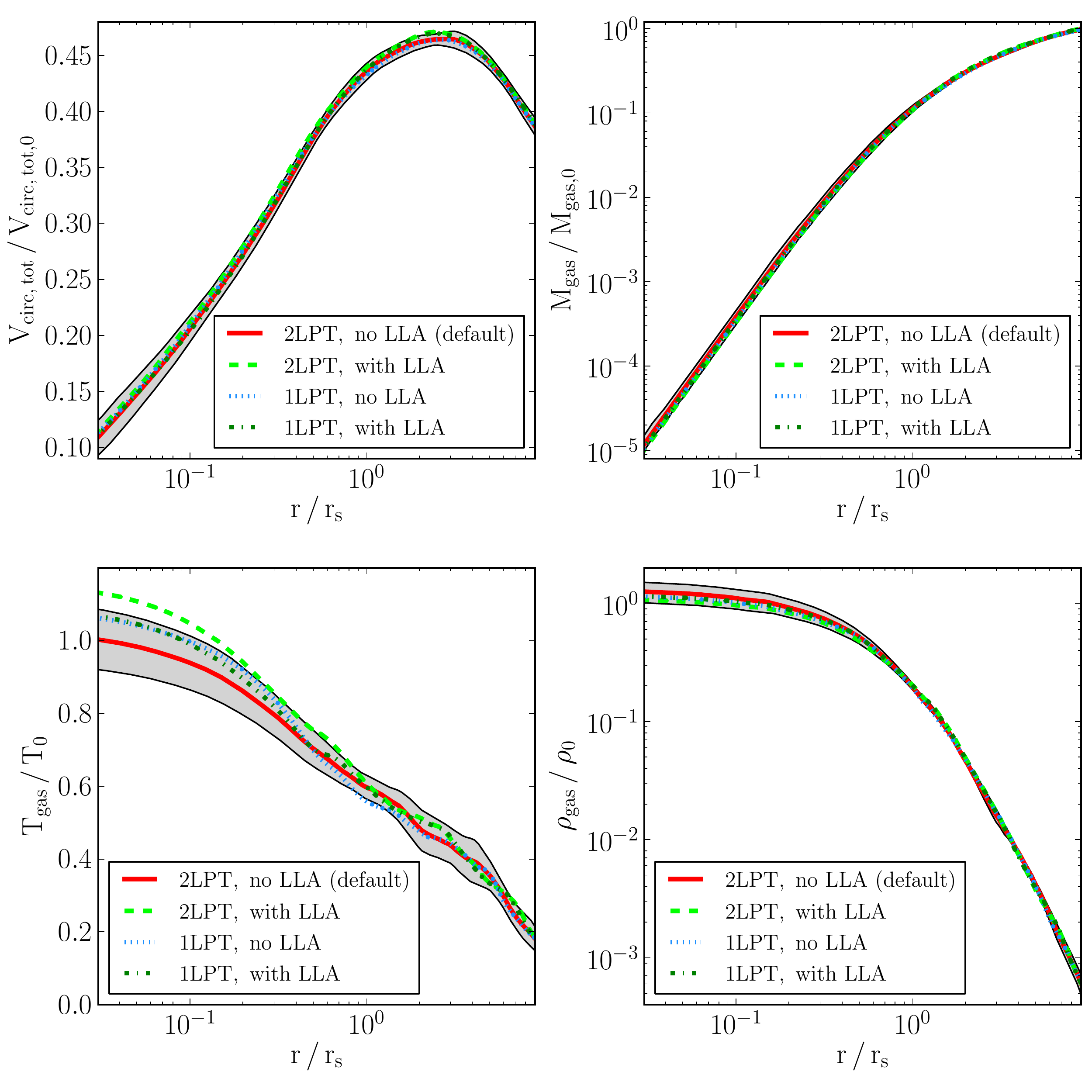}
\caption{The total circular velocity, gas mass, gas temperature and gas density profiles of halo 2 for the four different initial condition settings under consideration. 
For comparison the variance of the numerical mean of the 15 halos, as it is defined in Section ~\ref{results}, is also plotted as grey shaded area.}
\label{ICS_hydro}
\end{figure*}

The simulated profiles using the different initial conditions  (see Fig. ~\ref{ICS_hydro}) show only small deviations from the profile of the reference run for all four quantities. Please, note that deviations in the temperature profile appear larger, since it is plotted with linear scale, whereas the other plots use a logarithmic scale. Nevertheless, deviations remain always smaller than the grey shaded area indicating the variance in the corresponding profile, meaning that the details of the initial conditions generator do not play a role in this paper. 

\subsection{Effect of the hydrodynamics solver}

\begin{figure*}
\centering
\includegraphics[scale=0.7]{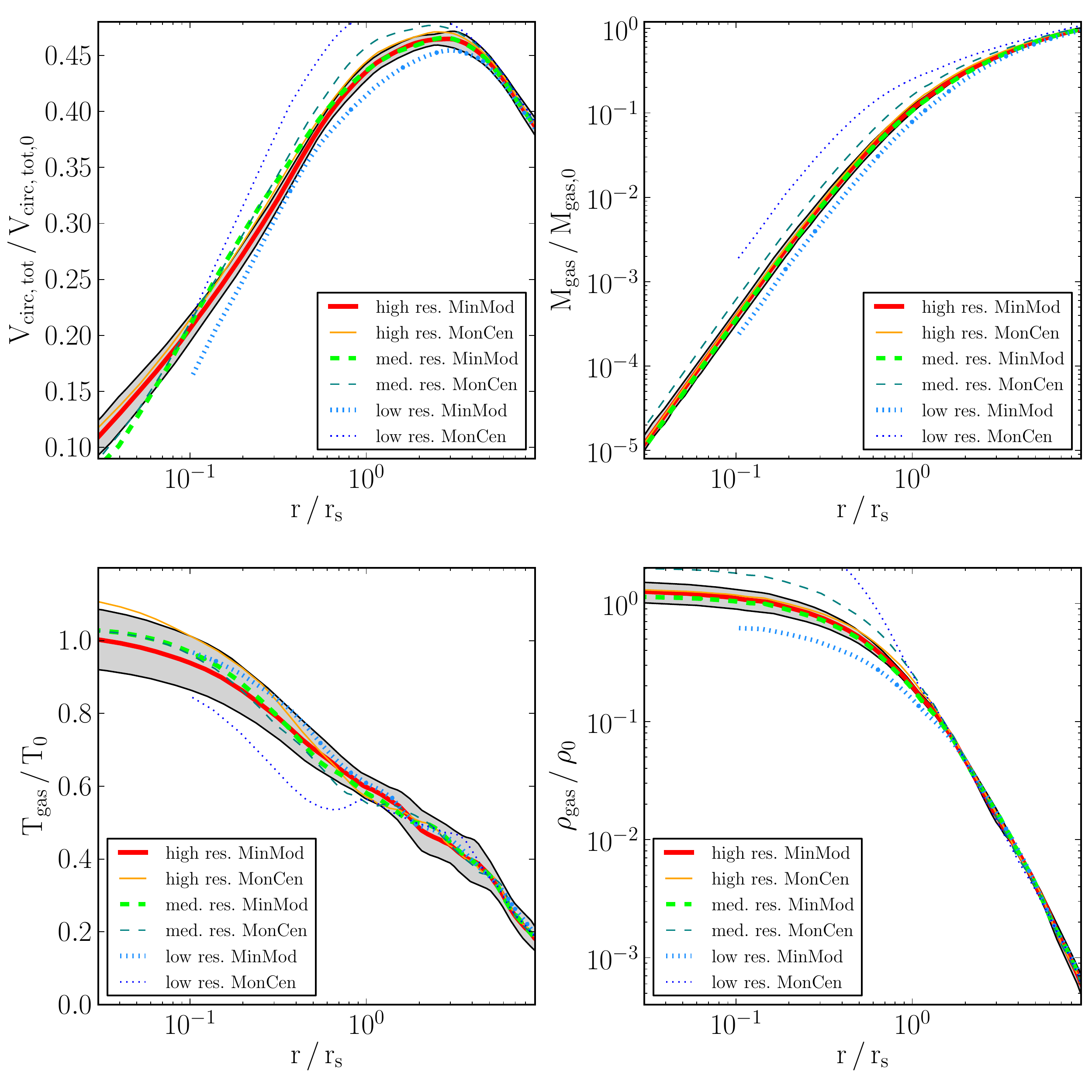}
\caption{The total circular velocity, gas mass, gas temperature and gas density profiles of halo 2 for the two different slope type hydro solver specifications, and respectively three different resolutions, under consideration. For comparison the variance of the numerical mean of the 15 halos, as it is defined in Section \ref{results}, is also plotted as grey shaded area.}
\label{ST_hydro}
\end{figure*} 
\begin{figure*}
\centering
\includegraphics[scale=0.7]{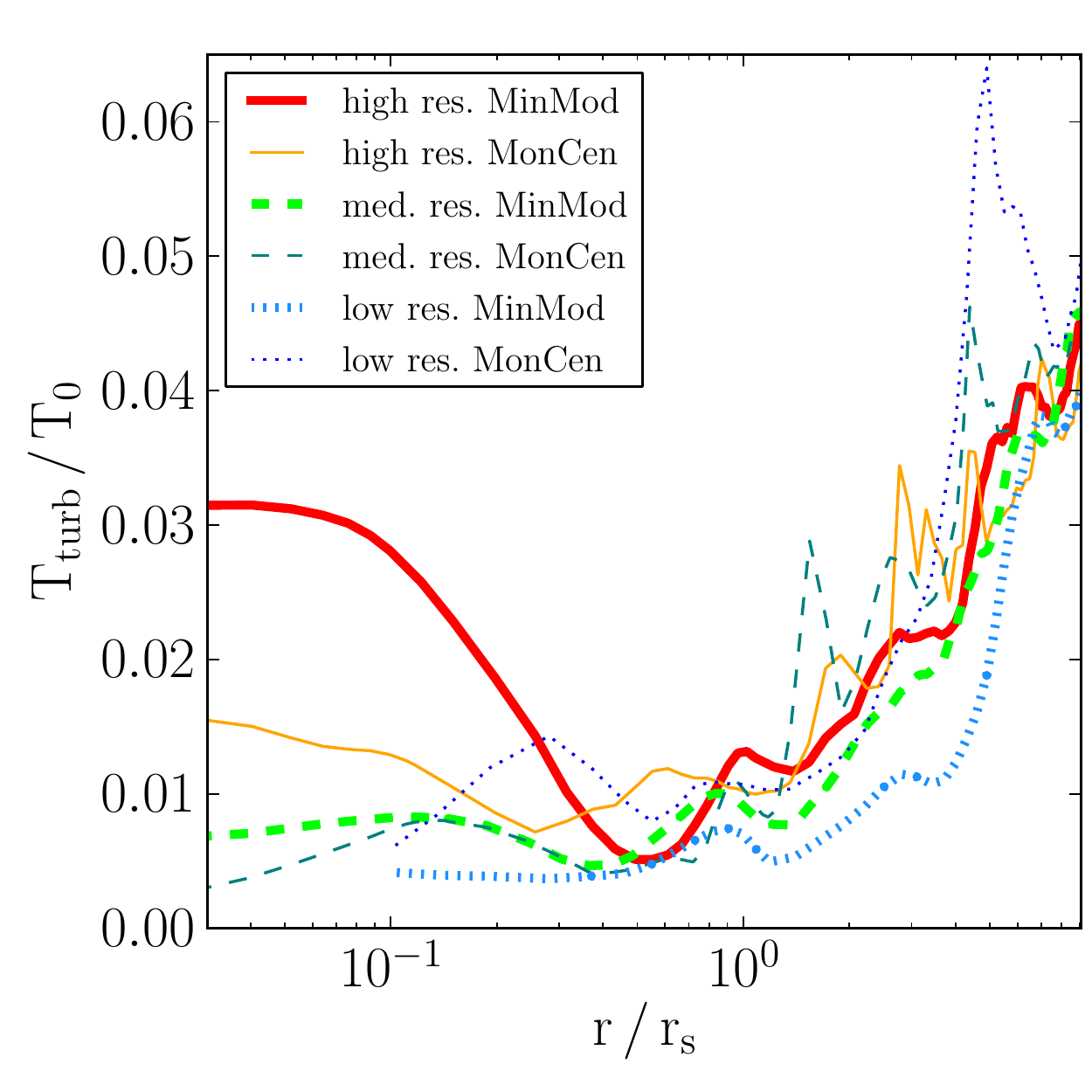}
\caption{The turbulent gas temperature of halo 2 for the two different slope type hydro solver specifications, and respectively three different resolutions, under consideration.}
\label{ST_turb}
\end{figure*} 
\begin{figure*}
\centering
\includegraphics[scale=0.7]{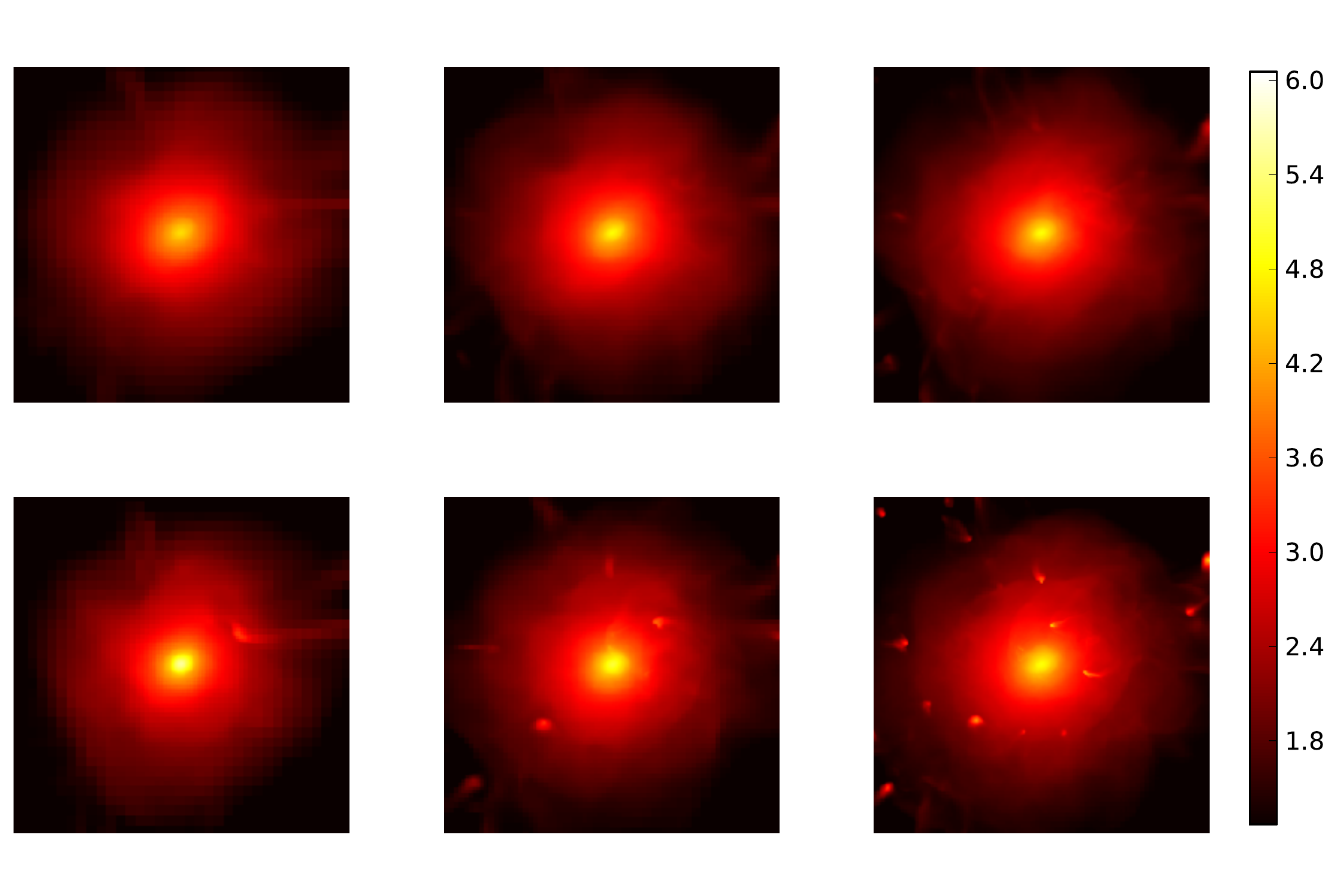}
\caption{Baryon density maps of halo 2 for the 2 slope type specifications: The upper 3 maps correspond to MinMod slope type and the lower three to MonCen slope type. The resolution is increasing from left to right: low ($\ell_{\rm ini}=9$), medium ($\ell_{\rm ini}=10$) and high ($\ell_{\rm ini}=11$). The colour bar unit is $\mathrm{log}_{10}(\rho_b/\bar{\rho}_b)$, where $\bar{\rho}_b$ is the baryon density of the universe. The side length of each map is $2r_{200}$, while the centre of the maps correspond to the centre of mass of the halo.}
\label{st_Density_Maps}
\end{figure*}  
\begin{figure*}
\centering
\includegraphics[scale=0.7]{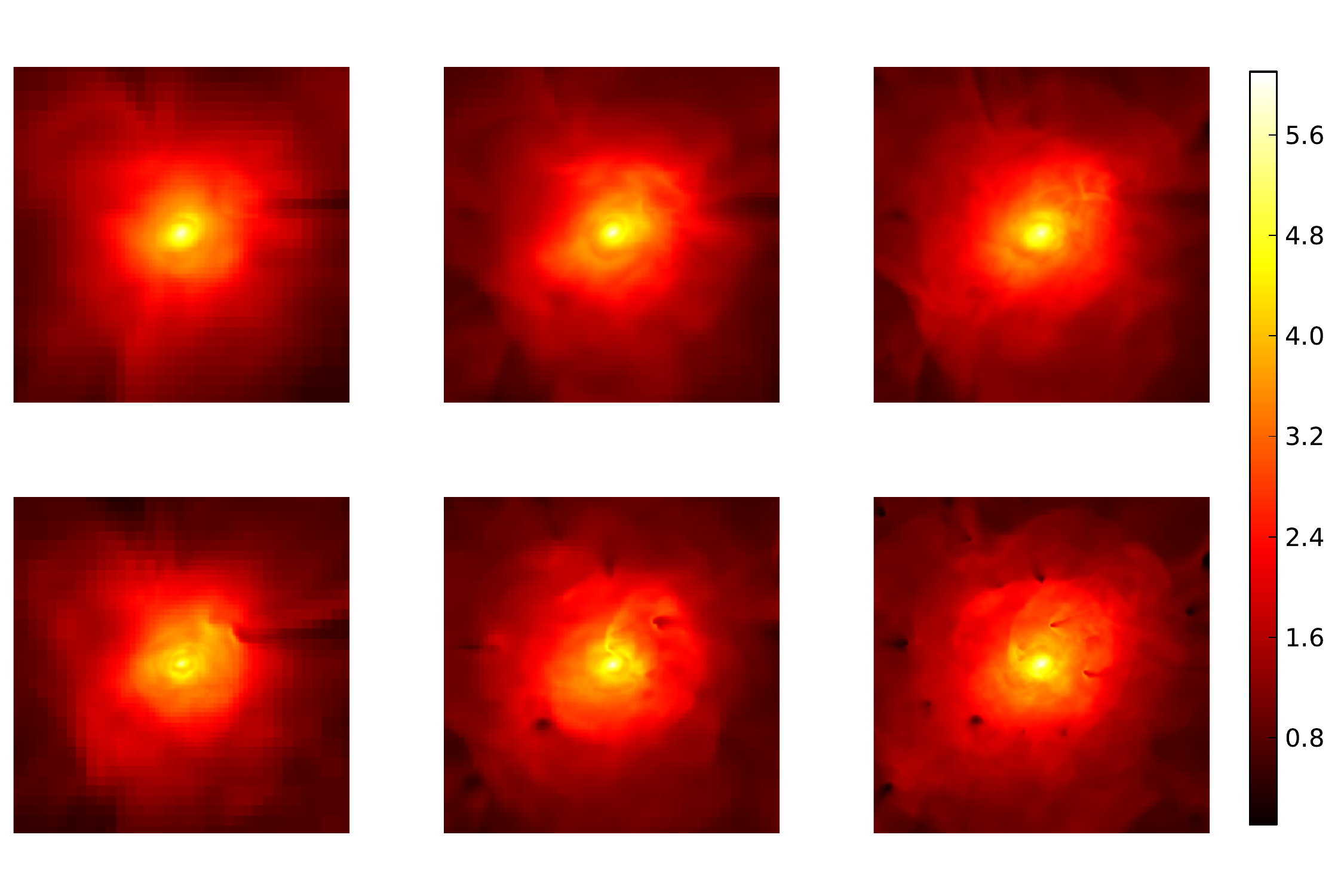}
\caption{Baryon thermal temperature maps of halo 2 for the 2 slope type specifications: The upper 3 maps correspond to MinMod slope type and the lower three to Moncen slope type. The resolution is increasing from left to right: low ($\ell_{\rm ini}=9$), medium ($\ell_{\rm ini}=10$) and high ($\ell_{\rm ini}=11$). The colour bar unit is $T_{gas}/T_{200}$, where $T_{200}$ is defined as $T_{200}=\frac{1}{3}\frac{m_p}{k_B}\frac{G\,M_{200}}{R_{200}}$. The side length of each map is $2r_{200}$, while the centre of the maps correspond to the centre of mass of the halo.}
\label{st_Temperature_Maps}
\end{figure*}   

We would like to test the robustness of our results with respect to the numerical parameters of the Godunov solver used in the RAMSES code.
As explained in \cite{2006A&A...457..371F}, these are the adopted Riemann solver, which can be either LLF or HLLC, and the slope limiter, either MinMod or MonCen.
For the Riemann solver, we have adopted the less diffusive one, the HLLC Riemann solver, and discarded completely LLF, because it is notoriously diffusive.
For the slope limiter, we explored the MinMod scheme, which is more diffusive but also more robust (our fiducial choice) and MonCen, which is more accurate but also less robust.

We show in Figure~\ref{st_Density_Maps} maps of the gas density distribution and in Figure~\ref{st_Temperature_Maps} maps of the temperature distribution for halo 2,
using our two slope limiters and our 3 different resolutions.
We see much more sub-structures for the more accurate, MonCen slope limiter, while for the more diffusive slope limiter MinMod, substructures seem to have been washed away, 
although, on closer look, they are still visible but very weak.
To quantify this spectacular effect, we have plotted in Figure~\ref{ST_hydro} the effect of the slope limiters on the measured profiles.
It appears now clearly that both slope limiters are converging to the same result, MonCen converging from above while MinMod is converging from below.
Interestingly enough, the a priori more accurate scheme systematically overestimate the mean converged profile, while the a priori more diffusive scheme
converges faster to the right solution, systematically underestimating the right answer. Our conclusion is therefore that once one uses enough resolution elements,
namely one million particles per halo, the influence of the slope limiter becomes smaller than the variance (shown in Fig.~\ref{ST_hydro} as the grey shaded area).

Since we have related the presence of substructure to the strength of turbulence, a very subtle effect, we would like to quantify the effect of the slope limiters to the 
level of turbulence in the halo. In Figure~\ref{st_Temperature_Maps}, one can see small and cold clumps with a clear bow shock structure ahead of them. 
This nicely resolved shocks could inject kinetic energy and power turbulence. We show in Figure~\ref{ST_turb} the turbulent temperature profile for our two slope limiters
and our 3 resolutions. One can see again that the more diffusive slope limiter gives us a smoother turbulence distribution, while the more accurate slope limiter preserves 
the substructure longer, giving rise to spikes in the turbulent energy in the vicinity of the substructure. 
Note that the overall profiles remain very similar, independently of the adopted resolution and slope limiter, especially if one considers the very large variance we observe in the
various turbulence profiles. Interestingly, our highest resolution temperature profiles show a rather large difference in the central region between MinMod and MonCen, the latter being colder than the former. 
A similar but opposite effect is seen in the turbulent temperature profile, proving that this is a minor transient feature. Indeed, at the exact time we have analysed our simulation's snapshot, turbulence was not entirely dissipated in the case of the MinMod slope limiter, while the MonCen slope limiter gives us a slightly more evolved snapshot for which kinetic energy has been transformed into heat.

\subsection{Summary}

We showed in this section, that changing the resolution to lower levels can have an effect onto the profiles of halo 2 mostly around 10\%.
The medium resolution profiles deviations, from the high res. profile, are far less than the variance coming from the individual halo nature. 
The low resolution profiles reach deviations from the high resolution ones, which are in the order of magnitude of the variance.

Altering the initial condition settings has only a minor effect onto the profiles of the four quantities considered, leading to deviations which are smaller than the extend of the individual halo variance, for almost all $r$-values.

The variation of the hydro solver slope limiter also causes deviations, which are smaller than the ones of the variance (of the order 10\%), but only when the highest resolution is applied. 
In case of the medium resolution the deviations are of the order of magnitude of the variance, and for the lowest resolution they are considerably larger. 
A very interesting feature of the hydro solver comparison plot (Fig. \ref{ST_hydro}) is that for MonCen the two lower resolution curves converge towards the high resolution one from above, whereas for the MinMod setting, they converge from below. 
In the latter case they also converge faster. 
The most noticeable feature however is that the MonCen runs show more substructure, and hence turbulence, than the runs with MinMod. This effect increases with increasing resolution.

With the analysis of this section we showed that generally the alteration of numerical parameters within the RAMSES code and the ICs provided by MUSIC can have a 10-20\% effect onto the profiles. 
A further step would be to quantify the effects, that the use of different codes would have, onto the profiles. 
This is however beyond the scope of this paper. Instead we estimate the variance coming from the use of different codes through the results of the first nIFTy paper \citep{2016MNRAS.457.4063S}: from the profile plots therein, it can be seen, that for the quantities relevant in this analysis the individual deviations in the subset of grid based and modern SPH codes, are of the order of 10-20\%, of the average. 
And mostly the agreement is better for the outer radii than for the inner ones.

\section{Conclusions}
\label{conclusions}

Novel methods in precision observational cosmology, like weak lensing and galaxy clustering, will enable observers to determine the matter power spectrum with high accuracy, 
down to relatively small scales, where non-linearities and baryonic effects will play an important role ($1\,h\,\mathrm{Mpc}^{-1} < k < 10\,h\,\mathrm{Mpc}^{-1}$). 
This will challenge theoreticians to compute the predicted power spectrum in this $k$-range with similar accuracy. 

In the present analysis, we have studied the internal structure of 16 galaxy group size halos with purely non-radiative hydrodynamics. 
The mass range was chosen because the mass distribution within groups will give the strongest contribution to the weak lensing signal. 
In addition, the scale free nature of non-radiative dynamics has the advantage that we can rescale our halo profiles, 
compute the average profiles and compare them with analytical predictions.    
By computing the 1-$\sigma$ standard deviation on the numerical mean profiles, we have found a variance of ~20\% for the most important gas quantities. 
This was interpreted as being due to different histories and internal dynamics of individual halos.
While this effect is of physical origin, changes in the numerical parameters on the other hand, lead also to measurable differences in the profiles, 
which are generally smaller than the variance. 
We conclude from this, that our simulations, with the highest resolution of $\Delta x_{min} \simeq$1~kpc and more than one million particles per halo, 
are accurate enough to reproduce the physics of galaxy group size halos  

In a further step, we have compared our numerical result to analytical profiles predicted by a classical theoretical model based on a polytropic gas in hydrostatic equilibrium within the NFW mass profile. 
We found an excellent agreement for the total circular velocity, with less than 10\% deviation between the numerical mean and the analytical profile.
With this result, we confirm that the NFW model is capable of describing the total mass distribution within halos, 
also when a collisional gas component is added to the dominant collisionless component, and that it provides a reliable analytical mass prediction for 
computing the power spectrum within the halo model approach \citep[see also][]{2008ApJ...672...19R}.  

For the thermodynamical properties of the gas component, we found a stronger disagreement between the analytical curve and the numerical mean. 
However, this deviation is generally smaller than 20\%, when the corrections due to the turbulent energy are taken into account, as we have done in our proposed corrected analytical model. 
Its main ingredient is a shallow but outwardly increasing turbulent temperature (equivalent to the specific turbulent energy profile). 
This behaviour was also found in the analytical analysis of \cite{2014MNRAS.442..521S} and \cite{2015MNRAS.448.1020S}, 
in which the authors derived their result from a set of differential equations describing the evolution of non-thermal, random motions in halos. 
A slightly different approach was used in \cite{2016arXiv160804388M}, who used what corresponds in our case to a constant turbulent temperature.     
Being able to understand better and describe more accurately the non-thermal contribution to the HSE, would be highly beneficial for mass estimates of galaxy groups 
and clusters based on X-ray observations, since these notoriously suffer from underestimation of the total pressure support (a problem known as the hydrostatic mass bias). 

Nevertheless, even after correcting the analytical model, we observe the strongest deviation from the numerical mean around $r=r_s$, 
consistently in all baryonic quantities. 
The gas entropy profile has provided us with an indication on the physical origin of this disagreement. 
We have indeed found two different thermodynamical regimes separated by the critical radius $r \simeq r_s$. 
In the central high density region, where merging of clumps and substructures leads to efficient mixing, 
we see a core of constant entropy. Here, isentropic evolution of an ideal gas with adiabatic index $\Gamma \simeq 5/3$ is recovered. 
In the outer part, however, we observe an atmosphere of decreasing density and increasing entropy, as predicted by our polytropic model with $\Gamma \simeq 1.2$. 

Hence the polytropic model of the baryonic component based on the HSE equation and the NFW distribution of the total mass, 
captures the essence of the dominant physical processes, but it also has its limitations: 
firstly, it does not take into account the additional non-thermal pressure support and secondly, it cannot predict the core of constant entropy we have observed. 
While for the first problem we have suggested an analytic correction, a detailed refactoring of the analytical model would be necessary to address the second issue, 
and is left for future work.   

\begin{figure}
\centering
\includegraphics[scale=0.575]{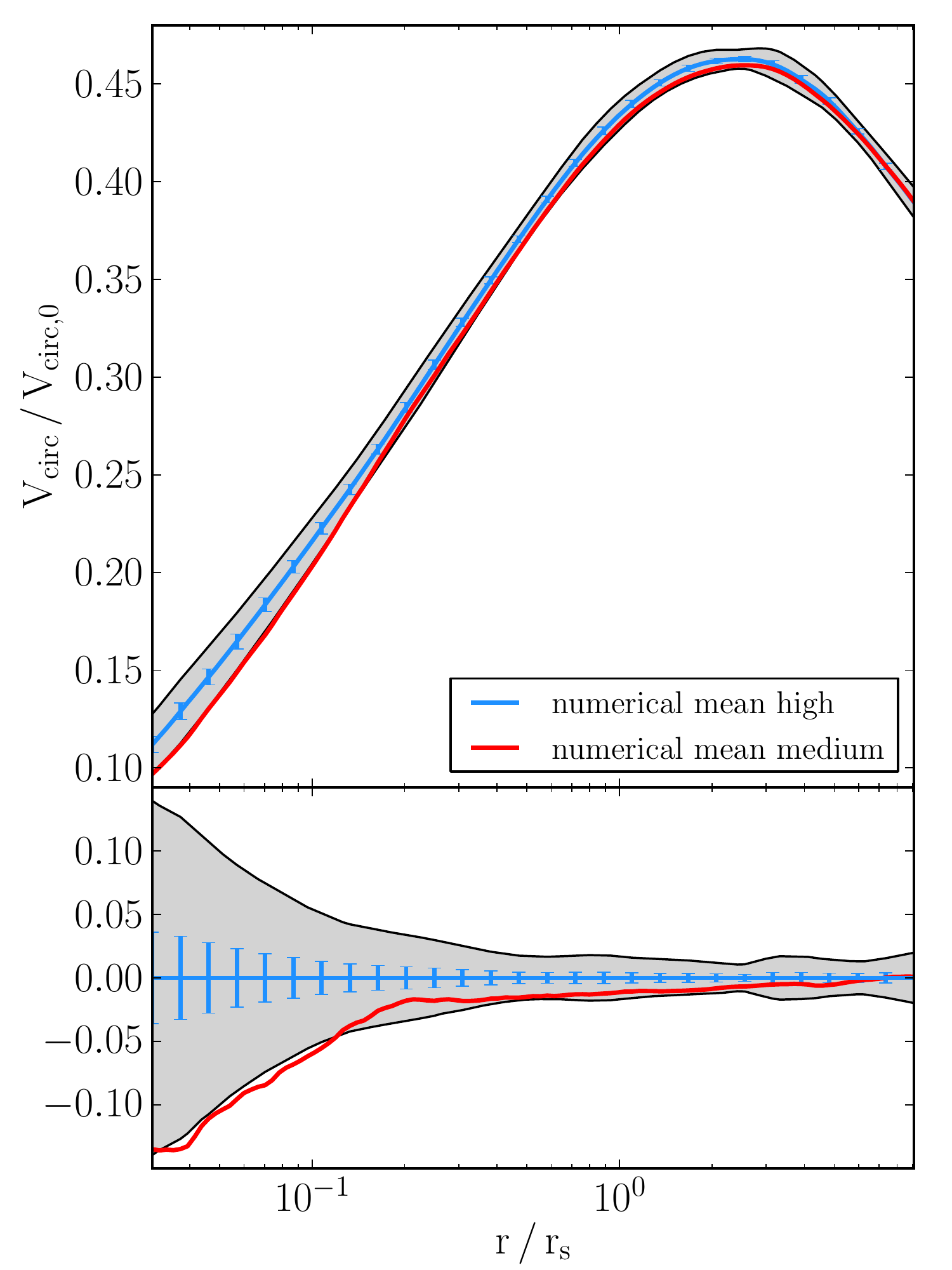}
\caption{Deviation between the high and medium resolution mean profile of the total circular velocity, the grey 1-$\sigma$ variance region and the blue error bars on the mean are calculated, as described in Section \ref{results}.}
\label{vcirc_mid_low}
\end{figure}

Furthermore, we have estimated the error onto our numerical mean profiles as the variance scaled with the inverse square root of the number of simulated halos. 
For the total circular velocity, we have found the error on the mean to be less than 1\% , for the range $r>0.2\,r_s$. 
This is good news for future projects on precision cosmology and the matter power spectrum. 
In the central region, the error increases however to 7\%. 
We conclude from this that the required 1\% precision is achieved within our non-radiative simulations down to very small scales. 
In order to estimate the possible bias due to resolution effects, we have calculated the numerical mean of the 15 halos also for the medium resolution case. 
For the circular velocity, we found that its deviation from the high resolution mean profile lies below 2\% for $r>0.2\,r_s$ (see Fig.~\ref{vcirc_mid_low}).
This indicates that our highest resolution results are converged within the estimated 1\% error bars.  

Overall, we can quantify the modification of the mass distribution due to the non-radiative baryonic gas component compared to pure dark matter simulations
using the main halo structural parameters $c$ and $r_{200}$. 
While for the N-body only runs we find as average for these parameters $c=11.7$ and $r_{200}=913\,\mathrm{kpc}$, 
in case of hydrodynamical runs we get $c=10.8$ and $r_{200}=991\,\mathrm{kpc}$. 
This indicates that, in the hydro simulations, halos are slightly more extended and less concentrated (see Table~\ref{Halo_Table}). 
We have found the decrease in the concentration parameter to be $\Delta c \simeq 1$ in average, 
which is significant even if this value is much smaller than the variance, with different halos having concentrations ranging $c=8$ to $c=18$. 

The idea to apply the NFW model to the total (gas + dark matter) mass, with different concentration parameters with or without a baryonic component, was introduced by \cite{2008ApJ...672...19R}.
These authors found almost no difference in the average $c$ value between their non-radiative and N-body simulations. 
However, their work is focused on the direct computation of the matter power spectrum, and hence they do not apply the zoom-in technique as we do, 
so that their mass resolution is much lower with $m_{DM} \simeq 10^9 \,M_{\odot}$. 
Accordingly, they only have 10000 particles per halo, in the considered halo mass range. Hence, our results are likely to be more accurate. 

Beyond non-radiative hydrodynamics, the inclusion of additional baryonic processes like cooling, star formation and feedback mechanisms is likely to produce stronger deviations into the mass profiles
especially in the centre, or increase the variance of the profiles even further.  
In a follow-up paper, we will explore these effects using various new physical processes on the same halo sample.

\section*{Acknowledgements}
The simulations for this work were performed on the zBox4 supercomputer at the University of Zurich. 
We thank S3IT for providing the infrastructure for this machine. 
We also thank the anonymous referee for greatly improving the quality of this paper, through constructive comments.

\bibliographystyle{mnras}
\def\apj{ApJ}
\def\apjl{ApJL}
\def\apjs{ApJS}
\def\aj{AJ}
\def\mnras{MNRAS}
\def\aap{A\&A}
\def\nat{nature}
\def\araa{ARAA}
\def\pasa{PASA}
\def\jcap{JCAP}
\bibliography{heap_without,romain}

\appendix

\section{Halo mass definitions}

\label{halo_mass_definitions}

Given the information about the halos which we obtain after running the HOP halo finder: the HOP halo masses and centres of mass of the halos, we can extract the halo profiles from the simulation outputs. 
With these data there are now three principle ways to find the halo mass $M_{200}$ and corresponding $r_{200}$ (defined with respect to the average mass density of the universe): \\\\

\textbf{Method 1} (HOP mass): use the HOP mass $M_{HOP}$ as $M_{200}$ and calculate $r_{200}$ from it via the relation:

\begin{equation}
M_{(<r_{200})}  = \frac{4 \pi}{3} \cdot \bar{\rho} \cdot 200  \cdot r_{200}^3
\end{equation}

\begin{eqnarray}
\Longrightarrow r_{200} & = & \sqrt[3]{\frac{3}{4 \pi} \cdot \frac{1}{\bar{\rho}} \cdot \frac{1}{200}  \cdot M_{(<r_{200})}} \\
\Longrightarrow r_{HOP}	& = & \sqrt[3]{\frac{3}{4 \pi} \cdot \frac{1}{\bar{\rho}} \cdot \frac{1}{200}  \cdot M_{HOP}}
\end{eqnarray}\\\\

\textbf{Method 2} (SOD meaning spherical overdensity): use the halo total mass profile $M_{(<r)}$ extracted from the simulation as function of $r$:

\begin{equation}
M_{(<r)} = \frac{4 \pi}{3} \cdot \bar{\rho} \cdot \Delta \cdot r^3
\end{equation}

\begin{equation}
\Longleftrightarrow \frac{M_{(<r)}}{r^3}  = \frac{4 \pi}{3} \cdot \bar{\rho} \cdot \Delta
\end{equation}

\begin{equation}
\Longleftrightarrow  \Delta  = \frac{3}{4 \pi} \cdot \frac{1}{\bar{\rho}} \cdot \frac{M_{(<r)}}{r^3}  
\label{Delta_equation} 
\end{equation}
 
When $\Delta=200$ one finds $r_{200}$ and $M_{(<r_{200})}$\\\\

\textbf{Method 3} (NFW mass profile): use the NFW mass profile $M^{NFW}_{(<r)}$ as function of $r$. The NFW parameters of the halo $\rho_s$ and $r_s$ need to be known (in our case we have extracted the from fitting the NFW circular velocity squared to the numerical profile:

\begin{equation}
M^{NFW}_{(<r)}  = 4 \pi \cdot \rho_s \cdot r_s^3 \cdot \left({\ln(1+r/r_s) - \frac{r/r_s}{1+r/r_s}}\right) 
\end{equation}

Inserting $\frac{M^{NFW}_{(<r)}}{r^3}$ as $\frac{M_{(<r)}}{r^3}$ into the expression for $\Delta$ \eqref{Delta_equation}:

\begin{eqnarray}
\Delta & = & \frac{3}{4 \pi} \cdot \frac{1}{\bar{\rho}} \cdot \frac{M^{NFW}_{(<r)}}{r^3} \\	
       & = & 3 \cdot \frac{\rho_s}{\bar{\rho}} \cdot  \left({\frac{r_s}{r}}\right)^3 \cdot \left({\ln(1+r/r_s) - \frac{r/r_s}{1+r/r_s}}\right) 		
\end{eqnarray}

Evaluating this expression at $r=r_{200}$ gives

\begin{eqnarray}
200 & = & 3 \cdot \frac{\rho_s}{\bar{\rho}} \cdot  \left({\frac{r_s}{r_{200}}}\right)^3 \nonumber \\ 
    &   &   \cdot \left({\ln(1+r_{200}/r_s) - \frac{r_{200}/r_s}{1+r_{200}/r_s}}\right)
\end{eqnarray}

\begin{figure}
\centering
\includegraphics[scale=0.575]{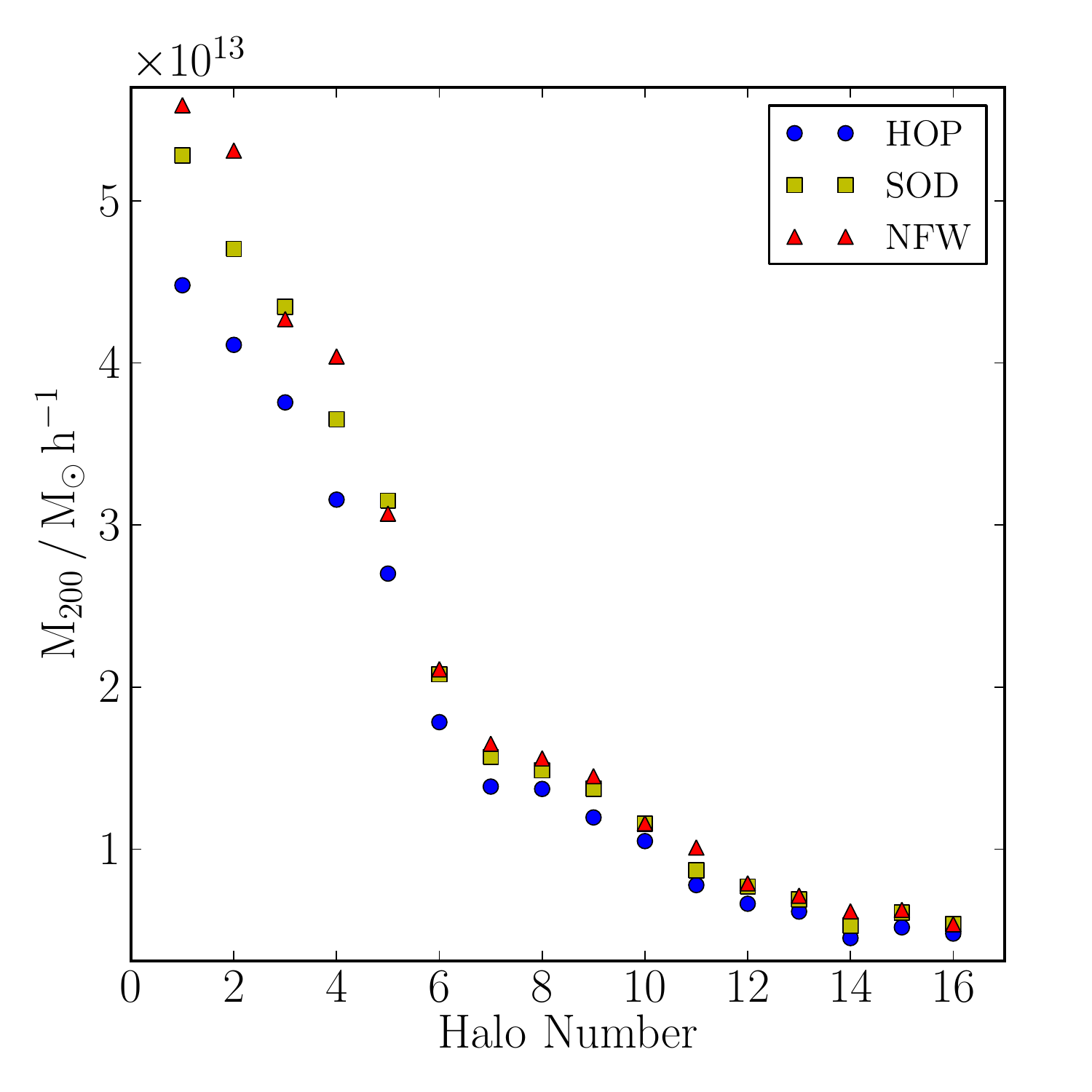}
\caption{Halo masses $M_{200}$ for the three different methods to obtain them: HOP (blue dots), SOD (yellow squares) , NFW (red triangles)}
\label{Halo_Mass_Comp}
\end{figure}

So one finds $r=r_{200}$ when $\Delta=200$ and can insert the value into $M^{NFW}_{(<r)}$ to obtain $M^{NFW}_{(<r_{200})}$.
\\\\
The resulting masses are displayed in Table \ref{Halo_Mass_Table} and Fig. \ref{Halo_Mass_Comp}. 
The average deviation between the NFW and SOD halo mass is 5\%, however no bias is observable. 
The HOP halo mass lies 13\% below SOD mass on average and a clear bias exists.

The HOP mass appears in our work because it is the one defining the halos, when we select our sample from the unigrid run. Though the spherical overdensity mass is the one which is used the most in literature, we give the NFW mass in Table \ref{Halo_Table} of the main text. The reason is simply that this mass corresponds to the other parameters which characterize our halos ($r_{200}$, $r_s$, $\rho_s$, $T_0$ etc.), as they are all connected through the assumption of hydrostatic equilibrium. Any other definition of $M_{200}$ would result in a slightly different value for $r_{200}$ and hence lead to a mismatch, when mixed with the other NFW parameters, as can be seen for example, in the definition of $c=r_{200}/r_s$.    

\begin{table}
\centering 
\begin{tabular}{c  c  c  c}
\\
\hline
Halo	& $M_{HOP}$ & $M^{SOD}_{200}$   & $M^{NFW}_{200}$   \\
	& (nbody)   & (hydro)       	& (hydro)	    \\
\hline
1		& 4.48  & 5.28 	& 5.59  \\	
2 		& 4.11  & 4.7  	& 5.31  \\
3 		& 3.76  & 4.35 	& 4.27  \\
4 		& 3.16  & 3.65 	& 4.04  \\
5		& 2.7   & 3.15 	& 3.07  \\
6		& 1.78  & 2.08 	& 2.11  \\
7		& 1.39  & 1.57 	& 1.65  \\
8		& 1.37  & 1.49 	& 1.56  \\
9		& 1.2   & 1.37 	& 1.46  \\	
10		& 1.05  & 1.16 	& 1.16  \\
11		& 0.778 & 0.869 & 1.01 	\\
12		& 0.663 & 0.768 & 0.789 \\
13		& 0.614 & 0.69 	& 0.713 \\
14		& 0.451 & 0.528 & 0.616 \\
15		& 0.517 & 0.608 & 0.625 \\
16		& 0.479 & 0.538 & 0.537 \\
\hline	
\end{tabular}
\caption[Halo Mass Definitions]{Comparison of halo mass definitions. All masses are stated in $10^{13} M_{\odot} h^{-1}$.}
\label{Halo_Mass_Table}
\end{table}

\section{Halo selection details}

\label{halo_selection_details}

In this appendix, we describe in more detail how we arrived at the 16 halos of our sample, from the initially 73'947 halos found in the considered mass range $ 5 \times 10^{12} M_{\odot} h^{-1} < M <  5 \times 10^{13} M_{\odot} h^{-1}$, in the unigrid run at $z=0$.

Since it is easier for the zoom-in simulations and the overall analysis process, that the halo candidates are relatively isolated, we applied the following isolation selection criterion onto the 73'947 found halos, with masses in the mass range of interest $ 5 \times 10^{12} M_{\odot} h^{-1} < M <  5 \times 10^{13} M_{\odot} h^{-1}$, in the first step: 
\begin{enumerate}
\item Check for each main halo (labelled with $1$), if it has another halo (labelled with $2$) within radius $r_{12} = a \left(r_{200,1}+r_{200,2}\right)$.
\item If this is the case, the mass ratios of the halo under consideration and the halos in its vicinity of radius $r_{12}$ are checked.
\item If $M_1 > b \cdot M_2$ then the main halo is kept in the sample, otherwise it is discarded completely. 
\end{enumerate}
As conservative minimum values of the selection parameters $a$ and $b$ we chose $(a,b)=(5,3)$. This criterion was fulfilled by 117 of the 73947 group sized halos. Among the other 73'830 halos were still extremely isolated ones, with a large value for either $a$ or $b$, e.g. $(a,b)=(13,1)$ for Halo 1 and Halo 14. Those cases were identified individually, so that another 26 halos were found and added to the 117 selected for the analysis. Table \ref{Halo_Selection_Table} shows the number of halos found for each parameter pair $(a,b)$.

Another step of selection was done by looking at the assembly history and the circular velocity curves of the 143 halos we have found so far: late mergers which had not yet accumulated 70\% of their final mass at $z=0.175$ were removed from the sample, as well as halos which showed atypical behaviour in their circular velocity plots. From the remaining 108 halos, we selected 16 which were lying close, above or below the average time evolution and circular velocity curves, to ensure that we have a sample representative of the typical behaviour of present day galaxy groups (see Table~\ref{Halo_Table} and Fig.~\ref{Time_Evolution_Plot} and \ref{Vcirc_Plot}). While as average for the time evolution selection we used the median of the 108 halos, for the average circular velocity curve we used the analytical expression from the NFW model \citep{1996ApJ...462..563N}, with the concentration parameter chosen as $c=9.6$ \citep{2011ApJ...740..102K}. Further we ensured, that the masses of the 16 selected halos represent the entire mass interval under consideration.

\begin{table*}
\centering 
\begin{tabular}{|c|c|c|c|c|c|c|c|c|c|c|c|c|c|}
\hline
$(a,b)$		& 1 & 2 & 3 & 4 & 5 & 6 & 7 & 8 & 9 & 10 & 11 & 12 & $\infty$ \\
\hline
1		& 883 & 874 & 862 & 854 & 845 & 834 & 831 & 822 & 819 & 813 & 805 & 802 & 195 \\	
2 		& 735 & 655 & 610 & 572 & 538 & 513 & 496 & 467 & 448 & 433 & 419 & 406 & \bf{21} \\
3 		& 552 & 436 & 360 & 316 & 273 & 252 & 233 & 213 & 193 & 180 & 173 & 163 & 1 \\
4 		& 404 & 273 & 222 & 179 & 146 & 131 & 114 & 95 & 85 & 79 & 74 & 66 & 0 \\
5		& 283 & 157 & \bf{117} & 83 & 65 & 56 & 49 & 42 & 34 & 31 & 29 & 26 & 0 \\
6		& 201 & 105 & 78 & 49 & 36 & 28 & 21 & 17 & 12 & 10 & 10 & 10 & 0 \\
7		& 133 & 64 & 44 & 23 & 17 & 14 & 13 & 10 & 9 & 7 & 7 & 7 & 0 \\
8		& 81 & 39 & 21 & 13 & 10 & 8 & 7 & 4 & 4 & 3 & 3 & 2 & 0 \\
9		& 50 & 23 & 10 & 5 & 4 & 3 & 3 & 3 & 3 & 2 & 2 & 1 & 0 \\	
10		& 30 & 13 & 5 & 3 & 2 & 1 & 1 & 0 & 0 & 0 & 0 & 0 & 0 \\
11		& 16 & 4 & 2 & 2 & 2 & 1 & 1 & 0 & 0 & 0 & 0 & 0 & 0 \\
12		& 11 & 4 & 2 & 2 & 2 & 1 & 1 & 0 & 0 & 0 & 0 & 0 & 0 \\
13		& \bf{6} & 1 & 1 & 0 & 0 & 0 & 0 & 0 & 0 & 0 & 0 & 0 & 0 \\
14		& 2 & 0 & 0 & 0 & 0 & 0 & 0 & 0 & 0 & 0 & 0 & 0 & 0 \\
15		& 1 & 0 & 0 & 0 & 0 & 0 & 0 & 0 & 0 & 0 & 0 & 0 & 0 \\
16		& 0 & 0 & 0 & 0 & 0 & 0 & 0 & 0 & 0 & 0 & 0 & 0 & 0 \\
17		& 0 & 0 & 0 & 0 & 0 & 0 & 0 & 0 & 0 & 0 & 0 & 0 & 0 \\
\hline	
\end{tabular}
\caption[Halo Isolation]{Halo isolation criteria table: number of halos found for each parameter pair $(a,b)$. The parameter $a$ increases vertically and the parameter $b$ horizontally.}
\label{Halo_Selection_Table}
\end{table*}

\end{document}